\documentclass[bibyear]{aa} 

\usepackage{xcolor}
\usepackage{graphicx}
\usepackage{txfonts}
\usepackage{booktabs}
\usepackage{siunitx}
\usepackage[colorlinks=true,citecolor=blue]{hyperref}
\usepackage{longtable}
\usepackage[flushleft]{threeparttable}
\usepackage{amsmath}
\usepackage{upgreek}
\usepackage{multicol}
\usepackage{multirow}
\usepackage{natbib}
\usepackage{orcidlink}
\usepackage{comment}
\usepackage{float}
\usepackage{txfonts}
\begin{document}

   \title{Colour evolution in the radio afterglow of GRB\,241025A}

   \author{S. Giarratana \inst{1}\fnmsep\thanks{email: stefano.giarratana@inaf.it} \orcidlink{0000-0002-2815-7291} \and O.~S. Salafia \inst{1,2} \orcidlink{0000-0003-4924-7322} \and L. Nava \inst{1} \orcidlink{0000-0001-5960-0455} \and G. Ghirlanda \inst{1,2} \orcidlink{0000-0001-5876-9259} \and M. Giroletti \inst{3} \orcidlink{0000-0002-8657-8852} \and S. Antier \inst{4} \orcidlink{0000-0002-7686-3334} \and M. Pillas \inst{5} \orcidlink{0000-0003-3224-2146} \and T. Hussenot-Desenonges \inst{4} \and A. Iskandar \inst{6,7} \orcidlink{0009-0003-9229-9942} \and M. Tanasan \inst{8} \and G. Oganesyan \inst{9,10} \orcidlink{0000-0001-9765-1552} \and N. Di Lalla \inst{11} \orcidlink{0000-0002-7574-1298} \and N. Omodei \inst{11} \orcidlink{0000-0002-5448-7577}}
 
    \institute{
    INAF Osservatorio Astronomico di Brera, Via E. Bianchi 46, I-23807 Merate, Italy    
    \and
    INFN -- sezione di Milano-Bicocca, Piazza della Scienza 3, I-20126 Milano (MI), Italy
    \and
    INAF Istituto di Radioastronomia, via Gobetti 101, 40129 Bologna, Italy
    \and
    IJCLab, Univ Paris-Saclay, CNRS/IN2P3, Orsay, France
    \and
    Sorbonne Université, CNRS, UMR 7095, Institut d’Astrophysique de Paris, 98 bis boulevard Arago, 75014 Paris, France
    \and
    Xinjiang Astronomical Observatory, Chinese Academy of Sciences, Urumqi, Xinjiang, 830011, China
    \and
    School of Astronomy and Space Science, University of Chinese Academy of Sciences, Beijing 100049, China
    \and
    National Astronomical Research Institute of Thailand (NARIT)
    \and
    Gran Sasso Science Institute, Viale F. Crispi 7, L’Aquila (AQ) I-67100, Italy
    \and
    INFN – Laboratori Nazionali del Gran Sasso, L’Aquila (AQ), I-67100, Italy
    \and
    W. W. Hansen Experimental Physics Laboratory, Kavli Institute for Particle Astrophysics and Cosmology, Department of Physics and SLAC National Accelerator Laboratory, Stanford University, Stanford, CA 94305, USA
    }

   \date{Received \dots; accepted \dots}
 
\abstract
{We present the observing campaign of the afterglow of GRB\,241025A, a $\gamma$-ray burst (GRB) whose prompt emission has been simultaneously detected by \textit{Swift}, \textit{Einstein Probe}, \textit{Fermi}/GBM, SVOM, Konus-\textit{Wind} and VZLUSAT-2 3U CubeSat. Our multi-wavelength campaign comprises radio, near-infrared, Optical and X-ray observations. The afterglow was clearly detected in all bands. We performed a semi-empirical fit of the data, showing that the afterglow behaviour can be reasonably reproduced by a single component, i.e. an ultra-relativistic shock. However, the results from the semi-empirical fit are inconsistent with the predicted evolution from the standard afterglow model in the slow cooling regime. Specifically, we found that at early times the synchrotron self-absorption frequency $\nu_{\mathrm{a}}$ should be at higher frequencies with respect to the ones sampled by our campaign, in order to explain the observed colour evolution in radio, namely the spectral evolution in time. To reconcile the prediction from the standard model with the observed data set, we fit the observations with a semi-analytical model, including a multiplicative factor $\tau_{\mathrm{enh}}$ to the optical depth which, in turn, artificially increases $\nu_{\mathrm{a}}$. We found that the radio colour evolution, together with the near-infrared, optical and X-ray emission, can be described reasonably well by a forward shock from a structured jet, provided that the optical depth in the shocked material is enhanced by a factor $\tau_{\mathrm{enh}} = 500$. We suggest that such enhancement in the optical depth can result from a population of cold electrons in the downstream material, i.e. electrons that were not accelerated by Fermi I process at the shock front, in agreement with the theoretical expectations previously reported in the literature. Overall, our work underscores the importance of systematic, multi-frequency, multi-epoch radio follow-ups of these extreme events.
}


\authorrunning{Giarratana, S. et al.}
\maketitle

\section{Introduction}
Radio observations of long-duration $\gamma$-ray bursts (GRBs) have proven to be crucial to study the evolution of the blast wave, its interaction with the medium surrounding the burst and the radiative processes that produce the observed afterglow emission. The radio band uniquely enables the tracking of the afterglow emission over its entire evolution: from the early phases ($\lesssim1$\,week post-burst), when the emission may be dominated by a reverse shock (RS; \citealt{Meszaros1997, Sari1999, Kobayashi2000, Gao2015, Resmi2016}), to later times (weeks, months or even years), when the forward shock prevails (FS; \citealt{Panaitescu2000}), all the way to the transition of the shocked material to a non-relativistic regime \citep{Frail2004, Berger2004, Frail2005}. High-cadence, multi-frequency observations can provide us with important information that might be otherwise missed. When included in multi-wavelength studies, this information can alleviate or even break the degeneracy in the multi-dimensional parameter space of the afterglow modelling, allowing us to derive the physical parameters that govern the emission. Despite its importance, only a small fraction of GRB afterglows have been observed in radio \citep{Chandra2012}. 
Moreover, only a handful of GRBs have been observed for long enough and over a wide frequency range to show an evolution of the radio spectrum in time. Such colour evolution in the radio emission can reveal the contribution of the RS \citep{Laskar2013, Perley2014, Laskar2016, Laskar2018b, Laskar2018c, Laskar2019, Rhodes2020, Rhodes2024} or constrain the passage of a synchrotron break frequency (from either a RS or a FS) within the band \citep{Laskar2018a, Perley2025}.

At 01:36:50 UTC on 2024 October 25, GRB\,241025A triggered the Burst Alert Telescope (BAT) onboard the Neil Gehrels \textit{Swift} satellite \citep{Ambrosi2024}. Located at RA(J2000)= 22:14:37.85; Dec(J2000) = $+$83\,d 34$^{\prime}$ 33.2$^{\prime\prime}$ \citep{Goad2024}, its prompt emission light curve showed a complex structure lasting approximately 180\,s \citep{Ambrosi2024}. The burst was also detected by the Gamma-ray Burst Monitor (GBM) onboard \textit{Fermi} \citep{Fermi2024}, the gamma-ray monitor (GRM) onboard the Space Variable Objects Monitor (SVOM; \citealt{SVOM-GRM2024}), the Wide--field X--ray Telescope (WXT) onboard the \textit{Einstein Probe} satellite (EP hereafter; \citealt{EP-GCN2024, EP-2024GCN2}), Konus-\textit{Wind} \citep{Konus2024} and the VZLUSAT-2 3U CubeSat \citep{VZLUSAT2024GCN}. A transient coincident with the burst was detected in the X-rays 116\,s after \citep{Ambrosi2024, Page2024GCN}, in the optical \citep{Jiang2024GCN, Pereyra2024GCN, SVOM-VT2024GCN, Watson2024GCN, Mohan2024GCN, Vinko2024GCN, Gupta2024GCN, Pankov2024GCN, Moskvitin2024GCN, Klingler2024GCN, Moskvitin2024GCNb, Rossi2024GCN} and in radio \citep{Giarratana2024GCN}. Optical spectroscopy between 3650--9500\,$\AA$ with the Nordic Optical Telescope (NOT) revealed a redshift $z=4.20$ \citep{NOT2024}, from which an isotropic equivalent energy in $\gamma$-rays of $E_\mathrm{\gamma,iso}=(5.5\pm0.5)\times10^{53}$\,erg was derived \citep{Konus2024}.

In this work we present our observations and interpretation of the GRB\,241025A afterglow emission. The structure of the paper is the following. In Section \ref{sec:observations} we present our multi-wavelength observations and the publicly available data. Empirical results from the monitoring campaign are presented in Section \ref{sec:results}. In Section \ref{sec:modelling} we apply a physically-motivated afterglow model to explain the observed emission of GRB\,241025A, and we discuss our results. Finally, we conclude with a summary in Section \ref{sec:conclusions}. Throughout the paper we assume a standard $\Lambda$--CDM model with $H_0 = 69.32$\,km\,Mpc$^{-1}$\,s$^{-1}$, $\Omega_m = 0.286$ and $\Omega_{\Lambda} = 0.714$ \citep{Hinshaw2013}.

\section{Observations and data analysis}
\label{sec:observations}
\subsection{Radio observations}
The Karl G. Jansky Very Large Array (VLA) observed the field of GRB\,241025A for seven epochs between 0.76 and 127 days post-burst (PI: Giarratana, project code: SF171028). Each observation was carried out at C (4 -- 8 GHz), X (8 -- 12 GHz) and Ku (12 -- 18 GHz) bands. The standard J2344$+$8226 was used as phase calibrator, while 3C286 was employed as band-pass and flux calibrator. The data were calibrated using the \texttt{CASA} pipeline and inspected for possible radio frequency interference. We then split the data at each frequency band in 2\,GHz chunks, in order to extract the maximum possible information from the radio data. The cleaned images were subsequently produced with the \textsc{tclean} task in \texttt{CASA} (Version 6.5.4., \citealt{McMullin2007}) using a Briggs weighting scheme with robust parameter set to $0.5$.
An unresolved source was detected at a position coincident with the X-ray and optical afterglow. To assess the flux density of the source, we selected a box with the \textsc{imview} task in \texttt{CASA}, centred on the peak of the emission, and we fitted the emission within the box with a Gaussian model (\textsc{imfit} task in \texttt{CASA}). Such fit includes the information on the r.m.s. noise level through the ``rms'' parameter of the task. In order to measure the r.m.s noise level of the map, we selected a large (of the order of tens to hundreds of times the size of the of synthesised beam) box close to the GRB position. The uncertainty on the flux density was computed as the sum in quadrature of the error on the Gaussian fit, which takes into account the r.m.s. noise level of the map, and a systematic error on the amplitude calibration, which we consider as 5\% of the total flux density of the source. For non-detections, we estimated the upper limit (given at a 3$\sigma$ confidence level) as three times the r.m.s. noise level of the map. We report the results of our VLA campaign in Table \ref{tab:grb25A_radio_log}.

\subsection{Optical and near-infrared observations and public data}
The Global Rapid Advanced Network Devoted to the Multi-messenger Addict (GRANDMA; \citealt{GRANDMA_2020}) started observing about 1 hour post-burst, with the Thai Robotic Telescope (TRT-SRO at Sierra Remote Observatory), the Tsinghua-Nanshan Optical Telescope (TNOT) at Nanshan Station of Xinjiang Astronomy Observatory, and the \href{http://kilonovacatcher.in2p3.fr/}{Kilonova-catcher} telescopes from the amateur community. The whole campaign was monitored with \textit{Skyportal} \citep{Coughlin2023}. We used the STDPipe pipeline \citep{Karpov2025} and its web interface\footnote{\url{https://grandma-stdpipe.ijclab.in2p3.fr}} for the analysis of our images. As for forced photometry, we compared the stars of our images to the Gaia DR3 catalog \citep{Gaia2023} for the $B$, $V$, $R$ (Optical) and $I$ (near-infrared; NIR) bands and to Pan-STARRS \citep{chambers2019panstarrs1surveys} DR1 images for the $r$ filter. We subtracted any constant flux contribution using either Pan-STARRS images taken previously or our own templates in the $R$ and $I$ filters. We also used the colour term mechanism in order to adjust the filter corrections between the templates and our campaign images, following the process described in \citet{Karpov2025}. We report the results of our campaign in Table\,\ref{tab:optical_data} and \ref{tab:NIR_data}. Throughout the paper, we consider $R$ and $r$ as the same (red) color, and similarly $I$ and $i$ are treated equivalently. 

Our data set was augmented by publicly available measurements reported via the General Coordinates Network\footnote{\url{https://gcn.nasa.gov/circulars/events/grb-241025a}}: an independent report from TRT-SRO \citep{Jiang2024GCN}; Colibri \citep{Watson2024GCN}, GROWTH-India \citep{Mohan2024GCN}; the RC80 robotic telescope at Piszkesteto Station of Konkoly Observatory \citep{Vinko2024GCN}; the LBT telescope \citep{Rossi2024GCN}. Measurements from the 1.3-m Devasthal Fast Optical Telescope (DFOT; \citealt{Gupta2024GCN}), the AZT-33IK telescope of Mondy observatory \citep{Pankov2024GCN} and the 1-m telescope of SAO RAS \citep{Moskvitin2024GCN, Moskvitin2024GCNb} were not included due to incomplete information on the exact mid-time or the magnitude system used. 

\subsection{X-ray observations and public data}

The Follow--up X--ray Telescope (FXT - \citealt{Zhao2024}) onboard the EP mission \citep{Weimin2022} observed the GRB afterglow in the 0.3-10 keV energy band starting $\sim$1 hour after the trigger. The FXT data analysis was performed with the Follow-up X-ray Telescope Data Analysis Software (\texttt{FXTDAS v1.2}) with the calibration database \texttt{CALDB v1.2}. The transient was detected in the first three observations (see Table~\ref{tab:FXT}). We extracted three spectra and analysed them with \texttt{Xspec v12.15.1}\footnote{\url{https://heasarc.gsfc.nasa.gov/docs/software/xspec/}}. For the spectral model we assumed a power-law  with Galactic absorption fixed to N$_{\mathrm{H,Gal}}$ = $1.21\times 10^{21}$ cm$^{-2}$ \citep{Kalberla2005, Willingale2013} and left free the intrinsic N$_{\mathrm{H,int}}$ but assumed it to be the same for the three spectra. Given the small number of counts in the third spectrum, we imposed it to have the same spectral index of the second spectrum. The results of the spectral fit, namely the de-absorbed flux integrated over the 0.3--10\,keV energy range and the photon index, are reported in Table~\ref{tab:FXT}. The corner plot of the spectral fit of three spectra is shown in Figure~\ref{fig:cornerFXT}. 

We obtained the \textit{Swift} X-Ray Telescope (XRT) unabsorbed flux light curve integrated in the 0.3--10\,keV energy range from the SWIFT BURST ANALYSER\footnote{\url{https://www.swift.ac.uk/burst_analyser/01262165/}} provided by the UK Swift Science Data Centre at the University of Leicester (UKSSDC, \citealt{Evans2007, Evans2009}). The absorption parameters are fixed to the values reported by the XRT website for this burst, namely N$_{\mathrm{H,Gal}}$ = $1.21\times 10^{21}$ cm$^{-2}$ \citep{Kalberla2005, Willingale2013} and N$_{\mathrm{H,int}}$ = $3.88\times 10^{22}$ cm$^{-2}$ for the Galactic and intrinsic host galaxy absorption, respectively. We note that the intrinsic absorption value reported by the XRT website is consistent with the one we obtained from the fit of the EP/FXT spectra. 

\section{Results}
\label{sec:results}

\subsection{Light curves}
\begin{figure*}
    \includegraphics[width=\textwidth]{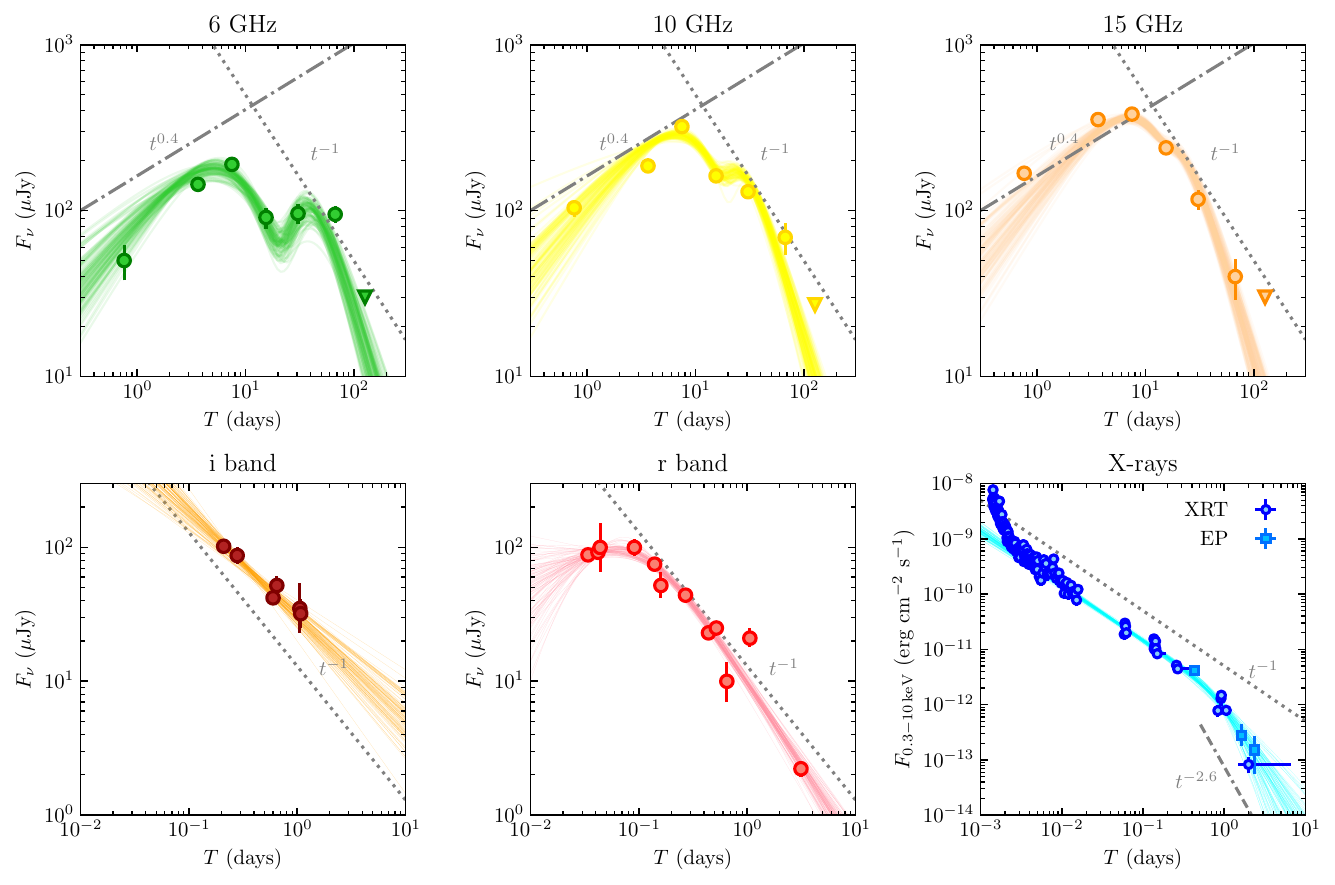}
    \caption{Multi-wavelength light curves of GRB\,241025A. Circles and squares with error bars in each panel show the flux density or flux of the GRB\,241025A afterglow in a specific passband (shown on top of the panel) at different times. Dot-dashed, dotted and double-dash-dotted gray lines show reference power law behaviors $t^{0.4}$, $t^{-1}$ and $t^{-2.6}$, respectively. In each panel, thin coloured lines show the light curves constructed from 100 random posterior samples of an empirical or semi-empirical model fitted to the observed evolution (see text).}
    \label{fig:semi-empirical-lc}
\end{figure*}
The radio light curves of GRB\,241025A at 6, 10 and 15\,GHz (respectively left, middle and right-hand panels in the top row of Figure\ \ref{fig:semi-empirical-lc}) exhibit a rise from our first observation (0.76\,d post-trigger) to the third one (7.5\,d post-burst) approximately described by $F_{\nu} \propto t^{0.4}$. The following behavior is initially chromatic (i.e.\ it differs among the frequencies) and not clearly described by a power law, until it transitions to a steep decay (steeper than $t^{-1}$) at $t\gtrsim 70$ d. We note here that, while our sampling cannot reveal short-timescale scintillation, the flux densities at each frequency appear to evolve smoothly in time, with no abrupt variability typical of this effect \citep{Rickett1990, Goodman1997, Walker1998, Alexander2019}. We further investigate the presence of scintillation in our data in Appendix \ref{appendix:scintillation}.

In order to quantitatively assess the  temporal decay of the NIR light curve (left panel, bottom row of Figure\ \ref{fig:semi-empirical-lc}), we fitted a power law model $F_{\nu} \propto t^{\alpha}$ to the data within a Bayesian framework, adopting a Gaussian likelihood and sampling the posterior probability through \texttt{DYNESTY} \citep{Speagle2020}. We obtained $\alpha=-0.77\pm0.10$ (median and 68\% credible interval). The thin orange lines in the figure show 100 model posterior samples for reference.

The $r$-band Optical light curve (middle panel, bottom row) shows a potential peak at $t\sim3\times10^3$\,s post-burst, followed by a decay. Given the apparent temporal structure, we fit a smoothly broken power law model to the Optical light curve,
\begin{equation}
    F_\nu = F_\mathrm{\nu,b}\left[\left(\frac{t}{t_\mathrm{b}}\right)^{-s\alpha_1}+\left(\frac{t}{t_\mathrm{b}}\right)^{-s\alpha_2}\right]^{-1/s},
\end{equation}
where $F_\mathrm{\nu,b}$, $t_\mathrm{b}$, $\alpha_1$, $\alpha_2$ and $s$ are free parameters. Uninformative, uniform-in-log priors on $F_{\nu,\mathrm{b}}$ and $t_{\mathrm{b}}$ and a uniform prior on $\alpha_1$ were adopted. The prior on $\alpha_2$ was uniform, with the requirement $\alpha_2<\alpha_1$, and the prior on the smoothness parameter $s$ was taken as uniform between $0.1$ (a smooth break) and $2$ (a very sharp break). The thin pink lines in the plot are constructed from 100 random posterior samples. The preferred break time is $t_\mathrm{b}=9.0_{-4.1}^{+5.7}\times10^{-2}\,\mathrm{d}$, with pre- and post-break slopes of $\alpha_1=0.70_{-0.56}^{+1.01}$ and $\alpha_2=-1.28_{-0.17}^{+0.10}$. We note that $\sim93\%$ of the posterior mass is at $\alpha_1 > 0$. The break time is therefore best described as a peak in the optical light curve, and the pre-peak slope shows that it can be interpreted as a deceleration peak, for which $\alpha_1\sim 2-3$. There is not substantial evidence for the presence of a second break (the log-Bayes factor in favour of a double smoothly broken power law is only $\ln K = 0.76\pm0.16$), but we note that, if present, the second break has a relatively well-defined time of $t_\mathrm{b,2}=2.8_{-1.3}^{+2.4}\,\mathrm{d}$, with pre- and post-break slopes of $\alpha_2=-0.97_{-0.18}^{+0.14}$ and $\alpha_3=-2.8_{-1.4}^{+0.9}$. 


Finally, the X-ray integrated flux light curve from \textit{Swift}/XRT and EP/FXT (right-hand panel, bottom row) features an initial steep decay (between $10^2$ and $10^3$\,s post burst) which we ascribe to the end of the prompt emission phase. 
The light curve after $10^3$ is better described by the smoothly broken power law model with respect to a single power law, with a log-Bayes factor $\Delta \ln K = 6.4\pm0.1$ that shows strong evidence in support of the presence of a temporal break at $t_\mathrm{b}=0.92^{+0.58}_{-0.30}\,\mathrm{days}$. The decay slopes before and after the break are $\alpha_1=-0.98\pm0.05$ and $\alpha_2=-2.6_{-1.10}^{+0.65}$. The thin cyan lines in the plot show 100 posterior samples of the model. The break time and the slopes are compatible with those of the potential break in the Optical discussed above. Therefore, we conclude that it is possible to interpret the X-ray break as an achromatic jet break.
We note that, between 0.1 and 1 day, the decay of all three light curves (NIR, Optical and X-rays) is consistent with $F_\nu\propto t^{-1}$ within three sigma.

\subsection{Spectra}
\begin{figure*}
    \includegraphics[width=\textwidth]{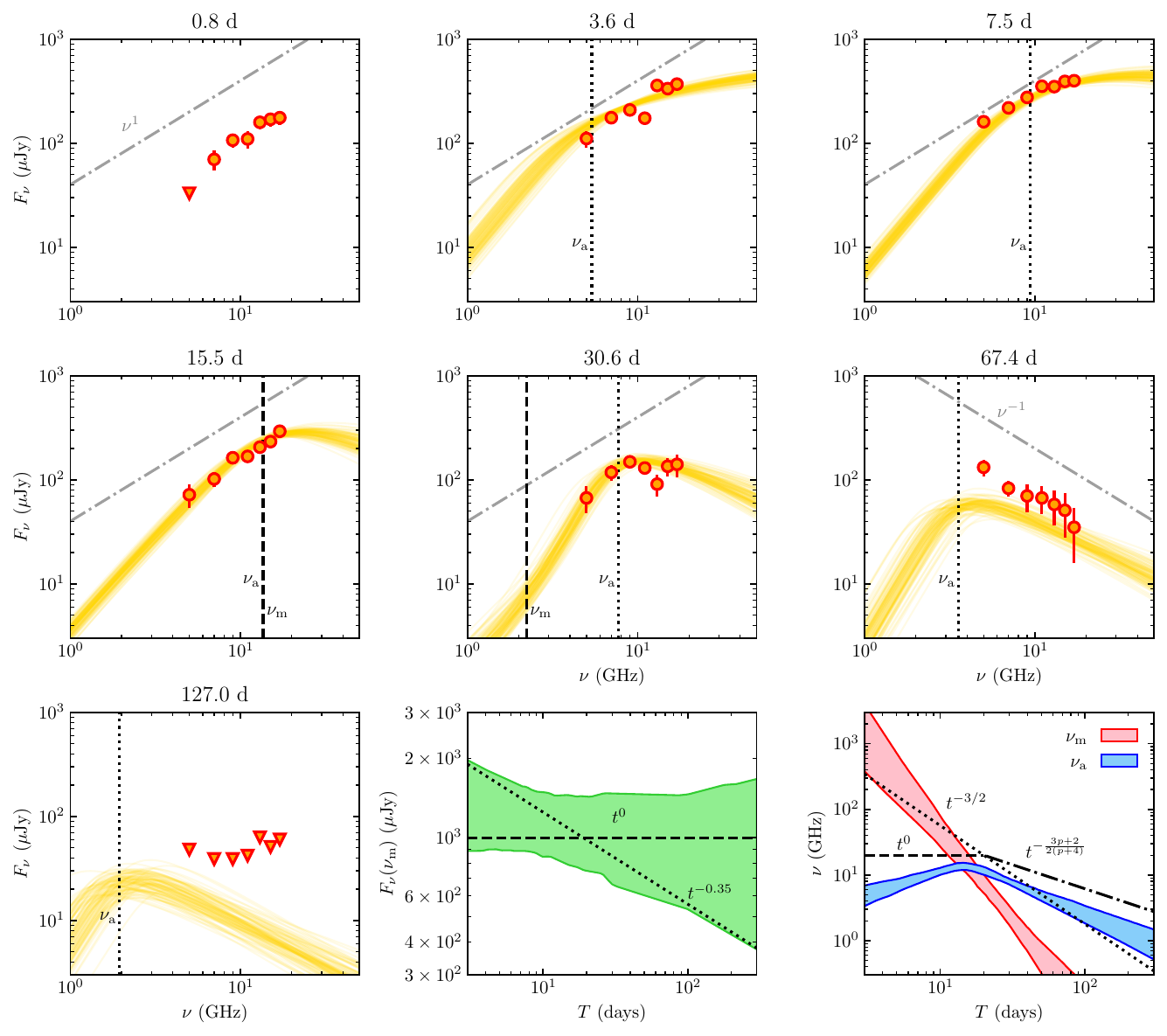}
    \caption{Radio spectra of GRB 241025A and synchrotron spectral parameter evolution. Orange circles with error bars in each panel show the flux density of the GRB 241025A afterglow at fixed times (shown on top of the panel) as measured by our radio observing campaign. Thin yellow lines show the radio spectra constructed from 100 random posterior samples of the semi-empirical model. The bottom-centre panel shows the 90\% credible region spanned by the time evolution of $F_\nu(\nu_\mathrm{m})$ (green band, 90\% credible region from the model), with a dashed line showing the expected evolution ($t^0$), and a dotted line showing $t^{-0.35}$ for reference. Similarly, the bottom-right panel shows the evolution of the break frequencies (red: $\nu_\mathrm{m}$; blue: $\nu_\mathrm{a}$). The dashed and dot-dashed lines show the expected evolution of $\nu_\mathrm{a}$ before and after the $\nu_\mathrm{m}$ crossing, respectively (the slope shown after the crossing assumes $p=2.5$); the dotted line shows the expected evolution of $\nu_\mathrm{m}$. All expected evolutions refer to the self-similar phase of a \citet{Blandford1976} blast wave expanding in medium with an ISM density profile \citep[e.g.][]{Granot2002}.}
    \label{fig:semi-empirical-spec}
\end{figure*}

The radio spectra of the first four epochs ($0.8\,\mathrm{d}< t < 15.5\,\mathrm{d}$) are relatively well described by a single power law $F_{\nu} \propto \nu^\beta$ with $\beta\sim 1$ (see Figure\ \ref{fig:semi-empirical-spec}, where grey dash-dotted lines guide the eye). Performing a formal fit of such a power law model to the first three spectra, we obtain $\beta=1.20_{-0.15}^{+0.20}$ (0.8 d), $\beta=1.0\pm0.1$ (3.5 d), $\beta=0.75\pm0.10$ (7.5 d) and $\beta=1.09\pm0.17$ (15.5 d). The spectrum at the fifth epoch ($30.6\,\mathrm{d}$) is convex, while the sixth epoch ($67.4\,\mathrm{d}$) shows a spectrum that decays with frequency, with $\beta=-0.9\pm0.3$. Within the synchrotron interpretation, the spectral behaviour is indicative of the passage of one or more break frequencies through the observed bands in the period $15.5\,\mathrm{d}\lesssim t \lesssim 67.4\,\mathrm{d}$.

The NIR-to-optical spectra at around 0.3--0.7 days post-burst can be described by a single power law with $F_{\nu} \propto \nu^{ -4}$. Since the extrapolation to the X-ray is clearly inconsistent with the X-ray observations at the same time (see Figure\,\ref{fig:LC_and_spectra}), we conclude that the NIR and optical data are affected by intrinsic absorption from the GRB host galaxy.
The X-ray spectra show evidence for a softening after $t\sim 1$ day, with the photon index transitioning from $\sim 1.8$ to $\sim 2-3$ (see Table\,\ref{tab:FXT} and Appendix \ref{appendix:butterfly}). 

\section{Interpretation}
\label{sec:modelling}

\subsection{The external shock scenario of GRB afterglow emission}
\label{subsec:modelling_general}

It is usually assumed that the afterglow phase of a GRB hails from the synchrotron emission of non-thermal electrons \citep{Paczynski1993, Panaitescu2000}. Specifically, the interaction between the jetted ultra-relativistic GRB ejecta and the cold surrounding medium triggers both a FS, which propagates into the circum-burst environment, and a RS, which compresses the GRB ejecta \citep{Rees1992, Meszaros1993, Meszaros2002}. Electrons at both the shock fronts are then accelerated up to relativistic energies by the Fermi I process. The process leads to a non-thermal distribution of electrons, which subsequently cool down via synchrotron and inverse Compton radiation. The spectrum at each time can be described by several power law segments which join at specific break frequencies that evolve with time \citep{Sari1998,Panaitescu2000, Granot2002}: (i) the self-absorption frequency $\nu_{\rm a}$, (ii) the injection frequency $\nu_{\rm m}$, and (iii) the cooling frequency $\nu_{\rm c}$. The physical and geometrical parameters of the standard model that describe the dynamics and the radiative processes are the kinetic energy $E_{\rm k}$; the initial bulk Lorentz factor $\Gamma_0$ of the GRB ejecta; the jet opening angle $\theta_{\rm j}$; the surrounding medium density $n(r) = A r^{-k}$, where $A$ is the normalisation, $r$ is the distance from the central engine and $k$ is the index describing the radial profile of the circum-burst medium density; the fraction of the shock downstream internal energy retained by the electrons $\epsilon_{\rm e}$ and by the magnetic field $\epsilon_{\rm B}$, the power law index $p$ of the energy distribution of the shock-accelerated electrons; the intrinsic host galaxy absorption $A_{V,\rm int}$. Regarding the circum-burst density profile, it is usually assumed that either $k=0$, which refers to a homogeneous interstellar medium (ISM hereafter) or $k=2$, which describes an environment with a wind-like density profile. All the parameters together determine the dynamics and the predicted observed radiation.

Concerning the overall multi-wavelength emission of GRB\,241025A, the afterglow light curves and spectra appear to be consistent with being produced by a single, dominant component (most likely, the FS). In fact, no evidence for the need of a second component (e.g., a RS in the optical or in the radio) is detected: the peaks of the optical and radio light curves occur too late, in the respective bands, to be attributed to the contribution of a RS. Throughout the paper, we therefore focus on one single emitting component.

\subsection{Deceleration peak and jet break}

As discussed above, the $r$-band light curve (bottom-left panel of Figure\ \ref{fig:semi-empirical-lc}) shows a peak at $t_{\rm pk} \sim 0.1$\,days post-burst, which can be ascribed to either the passage of $\nu_{\rm m}$ across the band, the deceleration peak \citep{Sari1999}, or both. 
Regarding the X-rays, the light curve steepening corresponding to the break observed in the light curve slope at $t_{\rm b} \sim 1$\,day is $\Delta\alpha=-1.6_{-1.1}^{+0.6}$. This cannot be ascribed to $\nu_{\rm c}$ passing through the band, because this would produce an insufficient steepening ($\Delta\alpha=-1/4$). A jet break, for which $\Delta\alpha$ can be as steep as $\sim -(3+p)/4$ (ISM case; less steep in the wind case), can explain the inferred steepening. 

If we interpret the above features as a deceleration peak and a jet break, we can obtain a first insight on the physical properties of the GRB~241025A jet. During the self-similar evolution of the blast wave \citep{Blandford1976}, the Lorentz factor of the region that dominates the observed flux evolves with the observer time as $\Gamma\propto t^{-(3-k)/2(4-k)}$. At the time of the peak, this Lorentz factor is approximately equal to the bulk Lorentz factor of the jet ejecta before deceleration, $\Gamma(t_\mathrm{pk})\sim \Gamma_0$, while at the time of the jet break the angular size of the relativistic beaming cone is comparable to the jet collimation angle, $\Gamma(t_\mathrm{j})\sim \theta_\mathrm{j}^{-1}$. Therefore,
\begin{equation}
\label{eq:gamma_theta}
    \Gamma_0\theta_\mathrm{j} \sim \left(\frac{t_\mathrm{j}}{t_\mathrm{pk}}\right)^{(3-k)/2(4-k)}.
\end{equation}
For $t_\mathrm{j}/t_\mathrm{pk}= 43_{-21}^{+37}$, this gives $\Gamma_0\theta_\mathrm{j}\sim 4.1_{-0.9}^{+1.1}$ in the ISM case, or $\Gamma_0\theta_\mathrm{j}\sim 2.6\pm0.4$ in the wind case. This relatively small product points to either a very small jet half-opening angle (union of 68\% credible intervals for wind and ISM cases: $\theta_\mathrm{j}\sim 1.2^\circ-3^\circ$ for a canonical $\Gamma_0=100$), a low jet bulk Lorentz factor ($\Gamma_0\sim 25-60$ for a canonical $\theta_\mathrm{j}=5^\circ$), or a combination of the two. As a caveat, we note that a small jet half-opening angle would cause the sideways expansion of the shock to be particularly fast \citep[see, e.g.,][]{vanEerten2012}, possibly leading to a transition of the shock dynamics from the self-similar deceleration phase to an exponential lateral expansion phase, which observationally could bring forward the jet break. 
Conversely, we note that, if the observed peak in the $R$-band were entirely due to $\nu_{\mathrm{m}}$ crossing the observing band, eq.\,\ref{eq:gamma_theta} would formally become an inequality, thereby providing a lower limit on the product $\Gamma_0\theta_j$.

\subsection{Self-similar phase}\label{sec:self_similar_phase}
The rough consistency between the decay rates of the NIR, optical and X-ray light curves between $0.1$ and $1$ day suggests that these bands are all on the same synchrotron spectral branch during this time interval. At around 0.1 d, the flux density ratio between $i$-band and 1 keV, which are separated by three orders of magnitude in frequency, is a bit less than two orders of magnitude. This suggests that the optical and X-rays are on the branch $\nu_\mathrm{m}\lesssim\nu<\nu_\mathrm{c}$ in a slow cooling scenario, or $\nu_\mathrm{c}\lesssim\nu<\nu_\mathrm{m}$ in a fast cooling scenario.
In fact, the slightly shallower slope in the $i$-band, though still consistent within $3\sigma$, may indicate that a spectral break has recently passed near this band. 
In the following, we assume a slow cooling regime, although we note that the fast cooling remains a possible alternative that is not investigated further in this work.
In the slow cooling scenario, the radio light curve indicates that $\nu_{\mathrm{a}}$ is close to our observing bands and therefore below the $I$-band. This is supported by the positive radio spectral slope at early times ($\leq30.6$\,days post-burst). This points to $\nu_\mathrm{m}\lesssim \mathrm{few}\times 10^{14}\,\mathrm{Hz}$ and $\nu_\mathrm{c}\gtrsim \mathrm{few}\times 10^{17}\,\mathrm{Hz}$ during this whole period. If the GRB is in a self-similar deceleration phase, the expected decay slope is $t^{-3(p-1)/4}$ in the ISM case, or $t^{-(3p-1)/4}$ in the wind case. For $t^{-1}$, this formally constrains $p\sim 2.3$ in the ISM case, or $p\sim 1.7$ in the wind case. Since $p<2$, the wind scenario is disfavoured under this interpretation. The wind case is also disfavoured following the evidence for a spectral softening in the X-ray data at around $1$ day (especially significant in our EP spectral analysis results, see Table\ \ref{tab:FXT}), which is consistent with $\nu_\mathrm{c}$ crossing the band while moving towards low frequencies -- a behaviour that is only expected in the ISM case.

\subsection{Radio evolution: $\nu_\mathrm{m}$ and $\nu_\mathrm{a}$ crossing the bands}
\label{sec:semi_emp_radio}

The observed radio spectral slope at $0.8\,\mathrm{d}<t<15.5\,\mathrm{d}$ is intermediate between the self-absorbed regime ($\nu<\nu_\mathrm{a}$) and the optically thin regime $\nu_\mathrm{a}<\nu<\nu_\mathrm{m}$. The spectral slope at $t=67.4\,\mathrm{d}$, on the other hand, is consistent with $\nu>\nu_\mathrm{m},\nu_\mathrm{a}$ (cfr.\ Figure\ \ref{fig:semi-empirical-spec}). This suggests that $\nu_\mathrm{a}$ is located within the observed frequency range during the first period, and that both $\nu_\mathrm{m}$ and $\nu_\mathrm{a}$ are located below that range at $t=67.4\,\mathrm{d}$. Since in general $\nu_\mathrm{m}$ decreases faster than $\nu_\mathrm{a}$, we expect that the two break frequencies cross each other while crossing the band. In order to test whether this description is consistent with the data, we set up a semi-empirical model as follows. We describe the optically thin synchrotron spectrum as
\begin{equation}
    F_{\nu,\rm thin} = F_{\nu_\mathrm{m}}\left[\left(\frac{\nu}{\nu_\mathrm{m}}\right)^{-s_1/3}+\left(\frac{\nu}{\nu_\mathrm{m}}\right)^{s_1(p-1)/2}\right]^{-1/s_1},
\end{equation}
where $F_{\nu_\mathrm{m}}$ is the flux density at $\nu_\mathrm{m}$, extrapolated at the meeting point of the two power law branches \citep{Granot2002}, and $s_1$ is a smoothing parameter. We describe the synchrotron self-absorption optical depth as \citep[e.g., eq.\, 52 from][]{Panaitescu2000}
\begin{equation}
\tau_{\nu} = \tau_{\nu_\mathrm{m}}\left[\left(\frac{\nu}{\nu_\mathrm{m}}\right)^{s_2 5/3}+\left(\frac{\nu}{\nu_\mathrm{m}}\right)^{s_2(p+4)/2}\right]^{-1/s_2},
\end{equation}
where $\tau_{\nu_\mathrm{m}}$ is the extrapolated optical depth at $\nu_\mathrm{m}$, and again $s_2$ is a smoothing parameter. Radiative transfer then gives the resulting spectrum as
\begin{equation}
    F_{\nu}(\nu) = F_{\nu,\mathrm{thin}}(\nu)\frac{1-\exp(-\tau_{\nu}(\nu))}{\tau_{\nu}(\nu)}.
\end{equation}
This description does not feature a discontinuous behaviour when $\nu_\mathrm{m}$ and $\nu_\mathrm{a}$ cross each other, contrarily to the commonly adopted recipe by \citet{Granot2002}. 

Focussing on times $t>1\,\mathrm{day}$ (i.e.\ after the putative jet-break), the time variation of the spectral break frequencies and of the normalization of the synchrotron spectrum can be described empirically by assuming power law evolutions for the parameters of the model, namely
\begin{equation}
\begin{split}
&F_{\nu_\mathrm{m}}(t) = F_{\nu_\mathrm{m},0}\left(\frac{t}{t_0}\right)^{\alpha_\mathrm{F}},\\
&\tau_{\nu_\mathrm{m}}(t) = \tau_{\nu_\mathrm{m},0}\left(\frac{t}{t_0}\right)^{\alpha_{\tau}},\,\text{and}\\
&\nu_\mathrm{m}(t) = \nu_{\mathrm{m},0}\left(\frac{t}{t_0}\right)^{\alpha_\mathrm{m}},
\end{split}
\end{equation}
where $t_0=10$ d is a reference time. We note that, defining $\nu_\mathrm{a}$ as the frequency for which $\tau_\nu=1$, we have $\nu_\mathrm{a}\propto t^{\alpha_\mathrm{m}+3\alpha_\tau/5}$ if $\nu_\mathrm{m}>\nu_\mathrm{a}$, and $\nu_\mathrm{a}\propto t^{\alpha_\mathrm{m}+2\alpha_\tau/(p+4)}$ otherwise .  We fitted the 9 model parameters $\lbrace s_1,s_2,F_{\nu_\mathrm{m},0},\tau_{\nu_\mathrm{m},0},\nu_{\mathrm{m},0},\alpha_\mathrm{F},\alpha_\tau,\alpha_\mathrm{m},p\rbrace$ to the radio data within a Bayesian approach, assuming wide,  poorly informative priors on all parameters, and sampling the posterior probability distribution with the Markov Chain Monte Carlo (MCMC) algorithm implemented in the \texttt{emcee} python package \citep{Foreman-Mackey2013}. We excluded the first epoch at $t=0.76\,\mathrm{d}$ from the fit, in order to avoid a possible change in the evolution regime at the putative jet break. The prior ranges and the results are summarized in Table \ref{tab:semi_emp_fit}. The corner plot is presented in Figure \,\ref{fig:semi_emp_corner} in Appendix \ref{appendix:semi-empirical}.

\begin{table*}[t]
    \caption{MCMC results for the evolving self-absorbed synchrotron spectrum model fitted to our radio data.}
    \centering\footnotesize
    \begin{tabular}{r|ccccccccc}
       Parameter & $s_1$ & $s_2$ & $\log(F_{\nu_\mathrm{m},0}/\mathrm{\mu Jy})$ & $\log(\tau_{\nu_\mathrm{m},0})$ & $\log(\nu_{\mathrm{m},0}/\mathrm{GHz})$ & $\alpha_\mathrm{F}$ & $\alpha_\mathrm{\tau}$ & $\alpha_\mathrm{m}$ & $p$  \\
       \hline
       Prior range$^{(a)}$ & $[0.5,3]$ & $[0.5,3]$ & $[1,6]$ & $[-3,10]$ & $[-3,3]$ & $[-4,0.5]$ & $[0,10]$ & $[-4,4]$ & $[2.01,3]$\\
       Result$^{(b)}$ & $0.63_{-0.11}^{+0.28}$ & $2.52_{-0.89}^{+0.45}$ & $2.95_{-0.24}^{+0.22}$ & $4.70_{-0.87}^{+0.97}$ & $-1.00_{-0.43}^{+0.32}$ & $-0.11_{-0.24}^{+0.19}$ & $5.71_{-0.95}^{+1.13}$ & $-2.67_{-0.59}^{+0.44}$ & $2.73_{-0.44}^{+0.25}$\\
       \hline
    \end{tabular}
    \\~\flushleft
    ~
    \tablefoottext{a}{Priors are uniform on the reported range when the parameters are expressed in the form shown in this table.}
    \tablefoottext{b}{The central value quoted represents the median of the marginalised posterior, while the error bars span the symmetric 90\% credible range.}
    \label{tab:semi_emp_fit}
\end{table*}

In each panel in the top row of Figure \ref{fig:semi-empirical-lc}, we show the modelling light curves, constructed from one hundred random posterior samples of the model, with thin coloured lines. Similarly, in Figure \ref{fig:semi-empirical-spec} we show the spectra at the corresponding time constructed from one hundred random posterior samples (thin yellow lines). In these panels, we also present the median of $\nu_\mathrm{a}$ ($\nu_\mathrm{m}$) at that time with a vertical dotted (dashed) line. In the last two panels of Figure \ref{fig:semi-empirical-spec}, shaded areas encompass the 90\% credible range (at any fixed time) of $F_{\nu_\mathrm{m}}$ (green band in the middle panel, bottom row), $\nu_\mathrm{m}$ and $\nu_\mathrm{a}$ (red and blue band, respectively, in the right-hand panel, bottom row -- the position of $\nu_{\mathrm{a}}$ from the model is defined as the frequency where $\tau_{\nu}=1$). The figures also show some reference power laws to guide the eye.

The figures demonstrate that the model produces a good description of the data, reinforcing the interpretation of the evolution as due to $\nu_\mathrm{m}$ and $\nu_\mathrm{a}$ crossing the band while also crossing each other. The evolution in the preferred model deviates somewhat from that expected in the self-similar phase ($\alpha_\mathrm{F}\approx -0.1_{-0.25}^{+0.2}$, which can accommodate the expected $\alpha_\mathrm{F}=0$, but shows a slight preference for a decaying evolution; $\alpha_\mathrm{m}\approx -2.7_{-0.6}^{+0.4}$ instead of $\alpha_\mathrm{m}=-1.5$): this could be ascribed to the effect of sideways expansion of the blast wave, given the late time of the observations. The model also shows a slight increase of $\nu_\mathrm{a}$ with time (the initial rising slope is formally $\alpha_\mathrm{m}+3\alpha_\tau/5=0.75_{-0.28}^{+0.21}$, even though this is only for the relatively short time between the first considered epoch at $3.6\,\mathrm{d}$ and the frequency crossing $\nu_\mathrm{m}=\nu_\mathrm{a}$ at $15.5\,\mathrm{d}$, after which $\nu_\mathrm{a}$ decays with a slope $\alpha_\mathrm{m}+2\alpha_\tau/(p+4)=-0.98_{-0.22}^{+0.17}$), and a preference for values of $p$ that are larger than the $p\sim 2.3$ inferred from the earlier optical and X-ray light curve evolution. Speculatively, these effects could be interpreted as the result of an electron acceleration efficiency that decreases with time, reflecting in a decrease of $\epsilon_\mathrm{e}$ and an increase of $p$. This could in principle lead to an increasing $\nu_\mathrm{a}$ and a steepening of the optically thin spectrum.
We note that a remarkably similar late-time, chromatic radio evolution is observed in GRB\,230815A by \citet{Leung2026}. Their description of the evolution, based on an even richer dataset, are similar to ours. This possibly points to a general feature of the late-time radio evolution of long GRB afterglows. 

\subsection{Failure of the standard afterglow model}
\label{sec:standard_afg_model_failure}
While the investigation in the previous section shows that we can successfully interpret the data with the described evolution of the break frequencies, the requirement that $\nu_\mathrm{a}\sim \mathrm{few}\,\mathrm{GHz}$ at around $1\,\mathrm{d}$  poses a challenge when compared with the constraints on the break frequencies discussed in Section \ref{sec:self_similar_phase}. To show this, we investigated the viable parameter space of a standard afterglow model, assuming an ISM density profile and a self-similar blast wave dynamics, when imposing the following constraints: $10^{14}\lesssim \nu_\mathrm{m}/\mathrm{Hz}\lesssim 5\times 10^{14}$ at $t=0.1\,\mathrm{d}$; $\nu_\mathrm{c}\gtrsim 10^{17}\,\mathrm{Hz}$ at $t=1\,\mathrm{d}$; $1\lesssim \nu_\mathrm{a}/\mathrm{GHz}\lesssim 15$ (the position of $\nu_\mathrm{a}$ is constant in the ISM case). We further required the flux density at 1\,keV ($=2.4\times10^{17}$\,Hz) to be within a factor of 2 from $1\,\upmu$Jy at $1.2\times10^{4}$\,s post-burst (from the results reported in Table\,\ref{tab:pre-break-fit}). These constraints were imposed on the model break frequency evolution of $\nu_\mathrm{m}$ and $\nu_\mathrm{a}$ as taken from \citet{Granot2002}; the cooling frequency $\nu_\mathrm{c}$ and the flux density $F_\nu(\nu=2.4\times 10^{17}\,\mathrm{Hz},t=1.2\times10^4\,\mathrm{s})$ were taken from the same work, but divided by $(1+Y)^2$ and $(1+Y)$, respectively, where $Y$ is the synchrotron self-Compton (SSC) parameter estimated\footnote{In practice, we set $1+Y$ equal to the maximum between the $x_\mathrm{SE}$ defined by \citet{Sari2001} and unity, to include both cases where SSC cooling is effective (the most common case in the relevant part of the parameter space), and those where it is negligible.} following \citet{Sari2001}. The parameter space was explored by generating samples uniformly distributed in the following intervals:
\begin{equation}
\begin{split}
    &\log(\epsilon_\mathrm{e})\in [-3,-0.5]\\
    &\log(\epsilon_\mathrm{B})\in [-7,-0.5]\\
    &\log(n/\mathrm{cm^{-3}})\in [-4,3]\\
    &\log(\eta_\mathrm{\gamma})\in [-2,-0.1]\\
    &p \in [2.2,3.5],
\end{split}
\end{equation}
where $\eta_\mathrm{\gamma}$ is the prompt emission efficiency, so that the blastwave isotropic-equivalent total energy is $E=E_\mathrm{\gamma,iso}(1-\eta_\gamma)/\eta_\mathrm{\gamma}$. A corner plot showing the distribution of samples that survive the cuts defined by the above constraints is shown in Figure \ref{fig:params_space_corner} in Appendix \ref{appendix:model_param_exploration}. The result shows that a standard afterglow model that includes SSC cooling cannot reproduce all the discussed constraints. In particular, the requirements on $\nu_\mathrm{m}$, $\nu_\mathrm{c}$ and $F_\mathrm{\nu,X}$ constrain the parameters in such a way that the self-absorption frequency cannot be higher than $1\,\mathrm{GHz}$ unless $p>3$. The latter condition, on the other hand, would make the light curve decay of such model inconsistent with the observed one, leading to a contradiction.

These considerations indicate that the self-absorption frequency at the time of our radio observations is too large to be in agreement with a standard afterglow model that accommodates the earlier NIR, Optical and X-ray evolution. As per the speculation in the previous section, this could hint at evolving microphysical parameters (as a result of an evolving efficiency of the electron acceleration process). Alternatively, or in addition to this, the self-absorption frequency could be enhanced by the presence of a population of relatively cold electrons (i.e.\ that do not participate in the acceleration process) in the shock downstream \citep{Eichler2005,Ressler2017, Warren2018, Warren2022}.

\subsection{Physical modelling of the afterglow phase}
\label{subsec:modelling_om_code}
\begin{figure*}
    \centering
    \includegraphics[width=0.95\textwidth]{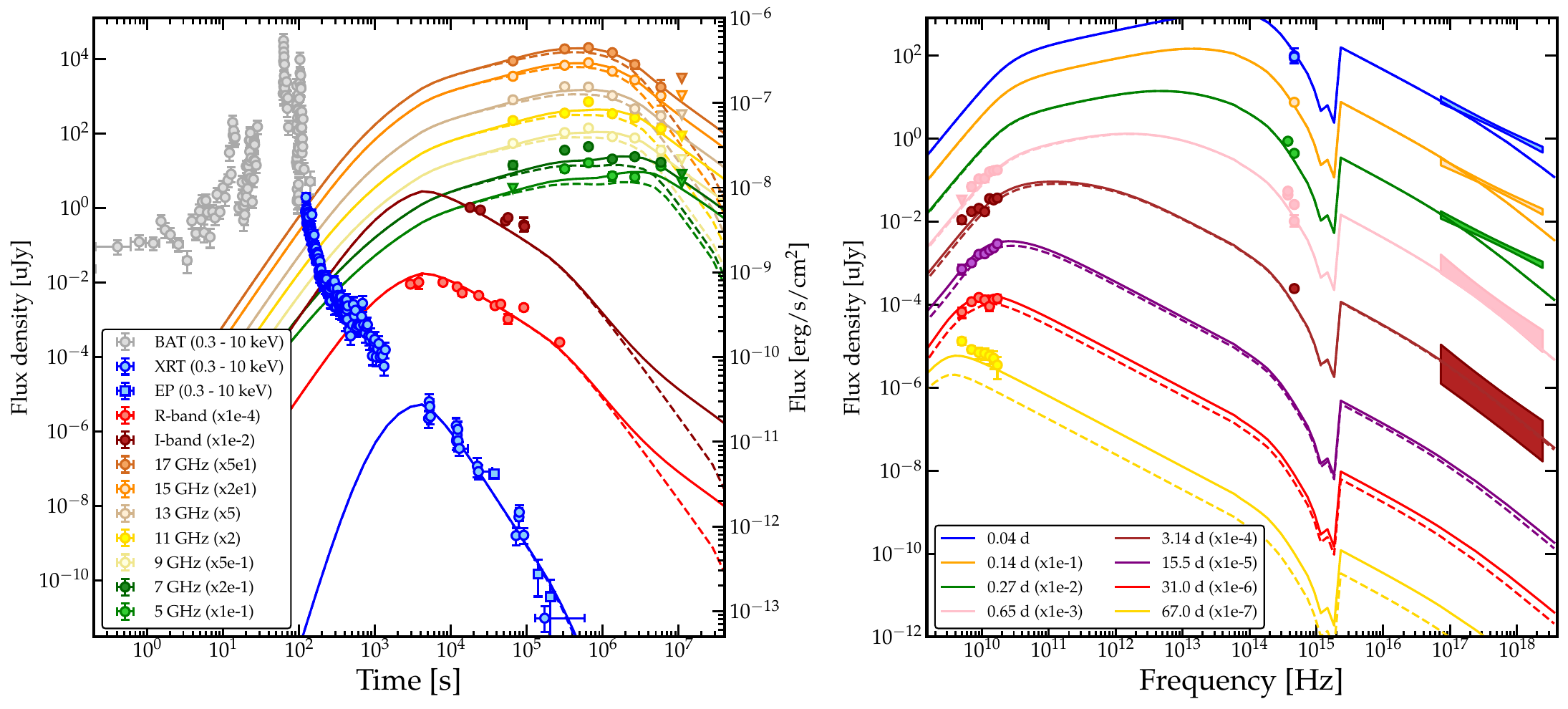}
    \caption{Observations and afterglow modelling of GRB\,241025A. \textit{Left panel}: multi-wavelength light curves. Prompt emission data from \textit{Swift}/BAT are shown as gray dots. The observed X-ray integrated flux (0.3--10\,keV) are represented with blue dots (\textit{Swift}/XRT) and squares (EP/FXT). Flux density measurements for the Optical and NIR are shown in red and brown, respectively. Radio flux density measurements are shown from the lowest band (5\,GHz, in green) to the highest band (17\,GHz, in light brown). \textit{Right panel}: spectra at different times. Radio, Optical, NIR and X-ray data are represented with the same colour for each epoch, from 0.04 days (blue) to 67 days (yellow) post-burst. The X-ray flux density spectra are presented as ``butterfly'' plots (see Appendix~\ref{appendix:butterfly} for a detailed description on the methods employed to calculate the butterfly plots). The afterglow model discussed in Section \ref{subsec:modelling_om_code} is represented with dashed (core of the jet) and solid (total emission, i.e. core + wings) lines.}
    \label{fig:LC_and_spectra}
\end{figure*}

In order to explain the entire multi-wavelength evolution of the GRB\,241025A afterglow in a physically-motivated fashion, we employ the model presented by \cite{Salafia2022}, which self-consistently computes the dynamics of the shocks from deceleration to the late side expansion phase. The model computes the resulting synchrotron emission accounting for the contribution from both the RS and the FS. The effects of inverse Compton scattering on the electron cooling, including Klein-Nishina corrections, are included. 
The shocked material is assumed to have a uniform velocity distribution within the jet opening angle (top-hat jet), and the viewing angle is along the symmetry axis. In order to model interstellar absorption in the host galaxy, we absorb the predicted Optical and NIR emission from the model with the \texttt{dust\_extinction} Python package \citep{Gordon2024}. 

As shown in the previous section, $\nu_\mathrm{a}$ must be forced to be at higher frequencies than those predicted by the standard afterglow model, in order to reconcile the observed multi-wavelength emission with the model prescriptions. This can be achieved by artificially increasing the synchrotron self-absorption optical depth of the shocked material. We explore this possibility by introducing an overall multiplicative factor $\tau_\mathrm{enh}$ to the optical depth computed within the model. Figure \ref{fig:LC_and_spectra}, dashed line, shows a set of light curves and spectra that attain a reasonably good description of the early (from approximately $2\times10^{3}$\,s to 31\,d post-burst) dataset, obtained by manually exploring the parameter space, and in particular by setting $\tau_\mathrm{enh} = 500$. Nevertheless, because of the jet break at $\sim 1\,\mathrm{d}$, the model underestimates the radio emission at 67\,d post-burst (see, e.g., the dashed line for the spectrum at this time in Figure\,\ref{fig:LC_and_spectra}). We find that an excess in the radio emission at this time cannot be attributed to an underlying supernova, as its associated specific luminosity $L_\nu\sim 3\times 10^{31}\,\mathrm{erg\,s^{-1}Hz^{-1}}$ is at least two orders of magnitude higher than that of the brightest known radio supernovae \citep{Bietenholz2021}.  As a way to account for the missing emission, we explored the possibility of an additional contribution from a wider jet component with a larger opening angle, but a lower isotropic-equivalent energy with respect to the component that dominates the emission at earlier times. The result is a structured jet with a narrow core and wider `wings'. We assume the physical parameters to be the same across the two components, changing only the initial isotropic-equivalent kinetic energy $E_{\mathrm{k}}$, the initial Lorentz factor $\Gamma_0$ and the half-opening angle $\theta_{\mathrm{j}}$. The model parameters are given in Table \ref{tab:model_params}. Subscripts `c' and `w' refer to the parameters of the narrow core and the wider jet, respectively. Solid lines in Figure\,\ref{fig:LC_and_spectra} show the total emission, while dashed lines single out the contribution from the narrow core. The model broadly explains the dataset from the early times up to very late times (127\,d post-burst), although the agreement with the X-ray slope is not perfect. The isotropic-equivalent kinetic energy profile of the two components roughly follows a power-law $E_\mathrm{k}(\theta) \propto \theta^{a}$ where $a\simeq-1$, a shallow trend that has been already proposed for GRB\,221009A \citep{OConnor2023, Gill2023}.
The low external density we find is not unprecedented (see, e.g., \citealt{Oganesyan2023} and references therein).
The total kinetic energy, corrected for the opening angles of the two jets, is $E \simeq 3\times10^{53}$\,erg. Such value would exceed the total available energy reservoir of a magnetar, ruling it out as possible central engine \citep{Thompson2004}.
Finally, we note that a late time excess in the radio emission could also be ascribed to an increase in the circum-burst density, or to late time energy injection from the central engine (see, e.g., \citealt{Leung2026}).

Overall, our work demonstrates that the entire evolution can be interpreted reasonably well within the standard afterglow scenario, provided that some (yet not fully specified) effect can enhance the self-absorption optical depth by a relatively large factor. We note that an even small number of relatively cold electrons can provide such a large enhancement, as found by \citet{Ressler2017}. The modelling light curves and spectra for the same choice of parameters but $\tau_{\mathrm{enh}}=1$ is presented in Figure\,\ref{fig:model_wo_enhancement}.

\begin{table}[]
\caption[]{Estimated physical parameters for the afterglow of GRB\,241025A.}
\centering
\begin{tabular}{lc}
\toprule
Parameter &Value\\ 
\midrule
$E_\mathrm{k,c}$ &$5.5 \times 10^{55}$\,erg\\
$\Gamma_{0, \mathrm{c}}$ &220\\
$\theta_\mathrm{j,c}$ &$1.7^{\circ}$\\
$n$ & $3 \times 10^{-2}$\,cm$^{-3}$\\
$\epsilon_\mathrm{e}$ &0.05\\
$\epsilon_\mathrm{B}$ &$2.5 \times 10^{-5}$\\
$p$ &2.7\\
$k$ &0\\
$A_{V,\mathrm{int}}$ &1.5\\
$\tau_\mathrm{enh}$ &500\\
\midrule
$E_\mathrm{k,w}$ &$5 \times 10^{54}$\,erg\\
$\Gamma_{0, \mathrm{w}}$ &50\\
$\theta_\mathrm{j,w}$ &$20^{\circ}$\\
\bottomrule
\end{tabular}
\tablefoot{The values refer to the two-component jet model with optical depth enhanced by a factor of 500. Subscript ``c'' refers to the parameters of the core jet, while subscript ``w'' refers to the parameters of the wider jet.}
\label{tab:model_params}
\end{table}

\section{Summary and conclusions}
\label{sec:conclusions}
In this paper, we presented the multi-wavelength observations of the afterglow of GRB\,241025A ($z=4.20$), whose prompt emission was simultaneously detected by \textit{Swift}, EP, \textit{Fermi}/GBM, SVOM, Konus-\textit{Wind} and the VZLUSAT-2 3U CubeSat. In particular, our radio observations between 5 and 17\,GHz allowed us to track the spectral evolution of the radio emission between 0.8 and 127\,d post-burst. The overall multi-wavelength afterglow could be described by a single evolving synchrotron component, i.e.\ a single shock, propagating through a circum-burst medium with an ISM-like density profile. Nevertheless, the results from our semi-empirical fit were inconsistent with the afterglow evolution predicted from the standard model. Specifically, the limits on the self-absorption frequency $\nu_{\mathrm{a}}$ inferred from the X-rays, Optical and NIR data at early times were not consistent with the actual position of the break frequency observed in the radio spectrum. In order to reconcile the radio emission with the standard model, $\nu_{\mathrm{a}}$ must be forced to be at higher frequencies with respect to those sampled by our observations.

We then fit the data with a semi-analytical afterglow model, introducing a multiplicative factor to the optical depth. We found that the emission could be described by a FS from a two-component jet, provided that the optical depth of the shocked material is artificially increased by a factor of 500. Such enhancement in the optical depth, which in turn forces $\nu_{\mathrm{a}}$ to higher values, could be explained by including a population of cold electrons, namely electrons in the shocked material that do not participate in the Fermi I acceleration process at the shock front. While the presence of a population of cold electrons is expected from the theory of ultra-relativistic shocks, a definite evidence of their contribution is still missing. In this paper, we suggested that detecting a radio colour evolution in other GRBs may represent a promising method to constrain the presence of cold electrons within the shocked material, although a theoretical effort is still needed in order to asses their true impact on the optical and radio emission. For GRB\,241025A, a detailed characterisation of the light curves and spectra has been possible due to a dedicated, multi-frequency, multi-epoch campaign - an approach typically reserved to the most interesting events. Therefore, our work highlighted the importance of tracking the radio afterglow of any GRB with a good temporal and frequency coverage. Only through consistent and systematic radio follow-up we can aim to understand the physical conditions driving the most energetic explosions in the Universe.

\begin{acknowledgements}
We thank the anonymous referee for their useful feedback.
The National Radio Astronomy Observatory is a facility of the National Science Foundation operated under cooperative agreement by Associated Universities, Inc. 
The observations presented in this work were carried out as part of project SF171028, approved in the framework of the Fermi - NRAO joint program agreement.
TNOT was sponsored by the Natural Science Foundation of Xinjiang Uygur Autonomous Region under No. 2024D01D32

The GRANDMA authors thank N. Khantanakorn, for TRT network, M. Freeberg, M. Serrau for the kilonova-catcher program, X. F. Wang, J. L. Liu, L. T. Wang, J. Mo, Y.S. Yan, A. Esamdin for TNOT, P. Gokuldass for VIRT, O. Pyshna for coordinations of the scientific exploitation,  H. Koehn for the scientific exploitation, S. Agayeva, E de Bruin, M. Masek, I. Abdi, D. Alk for the operations, M. Coughlin, D. Turpin, N. Guessoum, P. Hello, P-A. Duverne, S. Karpov, T. Pradier for the core team. GRANDMA is supported by the CNRS National Programme de Haute Energies.

SG acknowledges financial support from the Istituto Nazionale di Astrofisica (INAF) for the RAdio to TeV Transient Sources (RATTS) funding, project code: C53C22002020001. SG acknowledges financial support from the Istituto Nazionale di Astrofisica (INAF) for the ``Bando di Astrofisica Fondamentale 2024'', project number: 1.05.24.07.04. This work has been funded by the European Union-Next Generation EU, PRIN 2022 RFF M4C21.1 (202298J7KT - PEACE).

\end{acknowledgements}

\bibpunct{(}{)}{;}{a}{}{,}
\bibliographystyle{aa}
\bibliography{bibliography}

\clearpage

\begin{appendix}
\section{GRB 241025A: radio, optical and X-ray observations}
\label{app1}
\begin{figure*}
\centering
    \includegraphics[width=0.8\textwidth]{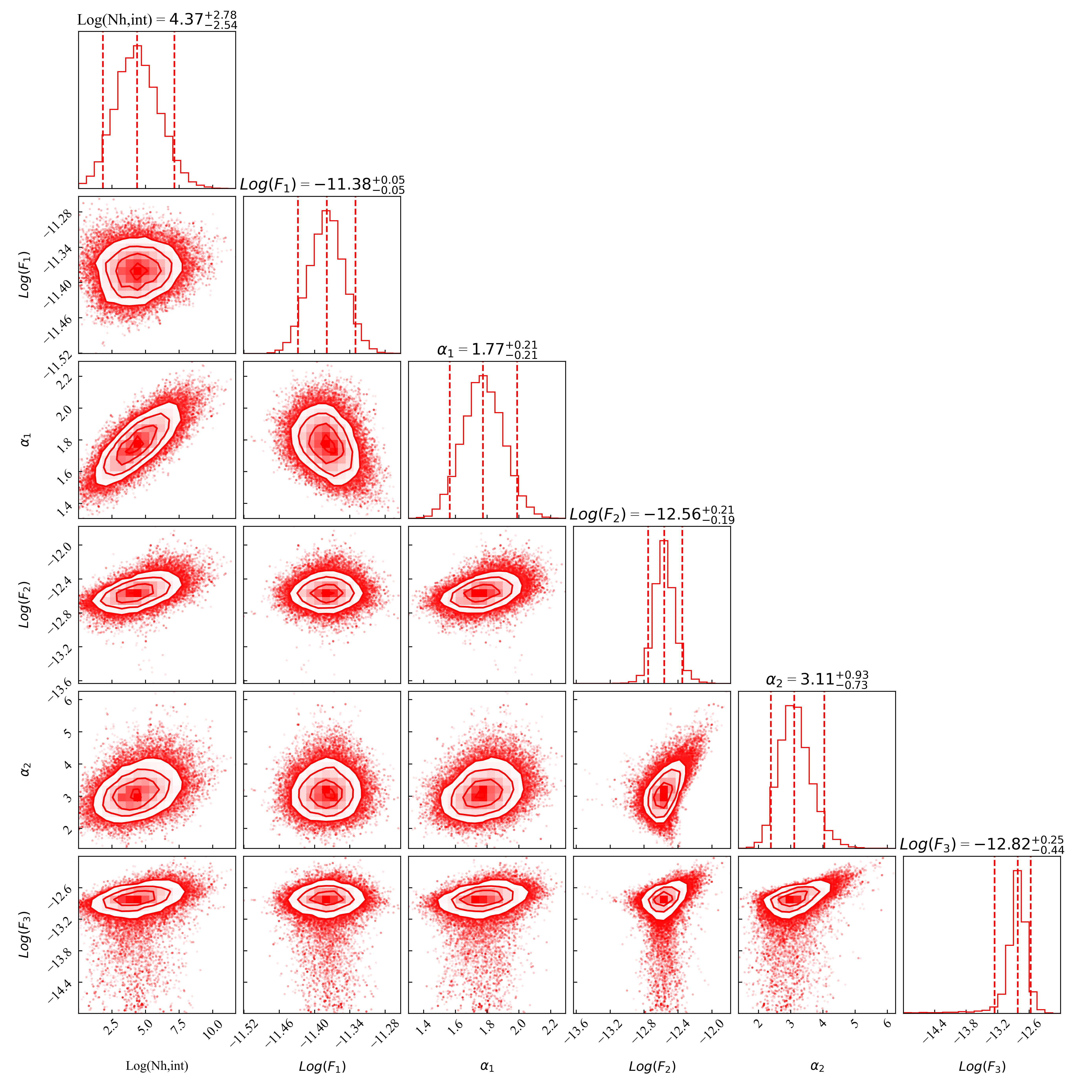}
    \caption{Corner plot of the spectral parameters obtained by the fit of the three EP/FXT observations. Subscripts 1,2,3 refer to the three observations reported in Table\ref{tab:FXT}.}
    \label{fig:cornerFXT}
\end{figure*}

In this appendix we report the log of our multi-wavelength campaign on GRB\,241025A. Radio observations with the VLA are presented in Table \ref{tab:grb25A_radio_log}. Optical and NIR observations and publicly available data are shown in Tables \ref{tab:optical_data} and \ref{tab:NIR_data}, respectively. Finally, X-ray observations with the EP/FXT are reported in Table \ref{tab:FXT}. Figure \ref{fig:cornerFXT} shows the corner plot of the spectral parameters obtained by the fit of the three observations. 

\begin{table*}
\small
\caption[]{Log table of our radio campaign on GRB\,241025A.}
\centering
\begin{tabular}{lcccccc|cccc}
\toprule
Date    &UTC    &T$_{m}$   &$\nu$ &$\Delta\nu$  &$F_{\nu}$ &r.m.s &$\nu$ &$\Delta\nu$  &$F_{\nu}$ &r.m.s\\ 
    &[hh:mm UT]   &[days]  &[GHz] &[GHz]  &[$\upmu$Jy] &[$\upmu$Jy beam$^{-1}$] &[GHz] &[GHz] &[$\upmu$Jy] &[$\upmu$Jy beam$^{-1}$]\\ 
\midrule
2024/10/25    &19:12 -- 20:42   &0.76   &6 &4 &$50\pm12$ &7 &5 &2 &$<33$ &11\\
 & & & & & & &7 &2 &$70\pm15$ &10\\
 & & &10 &4 &$104\pm13$ &8 &9 &2 &$107\pm17$ &10\\
 & & & & & & &11 &2 &$110\pm22$ &14\\
 & & &15 &6 &$168\pm16$ &8 &13 &2 &$159\pm23$ &13\\
 & & & & & & &15 &2 &$170\pm25$ &12\\
 & & & & & & &17 &2 &$176\pm26$ &15\\
\midrule
2024/10/28    &16:34 -- 18:04   &3.65   &6 &4  &$144\pm15$ &8 &5 &2 &$111\pm21$ &12\\
 & & & & & & &7 &2 &$176\pm18$ &10\\
 & & &10 &4 &$186\pm17$ &8 &9 &2 &$209\pm22$ &10\\
 & & & & & & &11 &2 &$174\pm22$ &12\\
 & & &15 &6 &$354\pm22$ &8 &13 &2 &$359\pm29$ &13\\
 & & & & & & &15 &2 &$335\pm28$ &13\\
 & & & & & & &17 &2 &$370\pm35$ &16\\
\midrule
2024/11/01    &13:10 -- 14:40   &7.51  &6 &4  &$190\pm18$ &8 &5 &2 &$161\pm23$ &12\\
 & & & & & & &7 &2 &$219\pm21$ &10\\
 & & &10 &4  &$321\pm21$ &8 &9 &2 &$276\pm23$ &11\\
 & & & & & & &11 &2 &$353\pm29$ &13\\
 & & &15 &6 &$382\pm24$ &9 &13 &2 &$350\pm29$ &14\\
 & & & & & & &15 &2 &$393\pm30$ &13\\
 & & & & & & &17 &2 &$399\pm35$ &18\\
\midrule
2024/11/09    &12:08 -- 13:38   &15.5   &6 &4  &$91\pm14$ &8 &5 &2 &$72\pm18$  &12\\
 & & & & & & &7 &2 &$102\pm17$  &10\\
 & & &10 &4 &$162\pm16$ &9 &9 &2 &$163\pm22$ &12\\
 & & & & & & &11 &2 &$168\pm21$ &13\\
 & & &15 &6 &$239\pm19$ &9 &13 &2 &$206\pm26$ &15\\
 & & & & & & &15 &2 &$233\pm26$ &15\\
 & & & & & & &17 &2 &$293\pm37$ &19\\
\midrule
2024/11/24    &14:57 -- 16:27   &30.6   &6 &4 &$96\pm14$ &7 &5 &2 &$67\pm19$  &11\\
 & & & & & & &7 &2 &$118\pm20$  &10\\
 & & &10 &4 &$130\pm14$ &8 &9 &2 &$149\pm20$ &10\\
 & & & & & & &11 &2 &$130\pm18$ &11\\
 & & &15 &6 &$117\pm17$ &9 &13 &2 &$91\pm21$ &14\\
 & & & & & & &15 &2 &$135\pm28$ &14\\
 & & & & & & &17 &2 &$141\pm35$ &17\\
\midrule
2024/12/31    &11:09 -- 12:39   &67.4   &6 &4 &$95\pm12$ &7 &5 &2 &$132\pm25$  &11\\
 & & & & & & &7 &2 &$83\pm15$  &10\\
 & & &10 &4 &$69\pm15$ &8 &9 &2 &$70\pm21$  &11\\
 & & & & & & &11 &2 &$67\pm20$  &12\\
 & & &15 &6 &$40\pm11$ &8 &13 &2 &$58\pm21$  &13\\
 & & & & & & &15 &2 &$51\pm23$  &13\\
 & & & & & & &17 &2 &$35\pm19$  &16\\
\midrule
2025/03/01    &05:39 -- 07:06   &127   &6 &4 &$<30$ &10 &5 &2 &$<48$  &16\\
 & & & & & & &7 &2 &$<39$  &13\\
 & & &10 &4 &$<27$ &9 &9 &2 &$<39$ &13\\
 & & & & & & &11 &2 &$<42$ &14\\
 & & &15 &6 &$<30$ &10 &13 &2 &$<63$ &21\\
 & & & & & & &15 &2 &$<51$ &17\\
 & & & & & & &17 &2 &$<60$ &20\\
\bottomrule
\end{tabular}
\tablefoot{T$_m$ is the time interval from the \textit{Swift}/BAT trigger to half of the observation. The uncertainty on the flux density takes into account the statistical error from the Gaussian fit, the r.m.s. noise level of the map and a systematic error on the amplitude calibration, which we consider as 5\% of the total flux density of the source. Upper limits are taken at a $3\sigma$ confidence level.}
\label{tab:grb25A_radio_log}
\end{table*}

\begin{table}
    \centering
    \caption{Log table of our $R$-band and $r$-band observations and publicly available data of GRB\,241025A.}
    \resizebox{\columnwidth}{!}{
    \begin{tabular}{lccccll}
    \hline
     Date & UTC  &T$_m$ &Mag &$F_{\nu}$ &Instrument &Ref\\
          &[hh:mm UT] &[days] &[AB] &[$\upmu$Jy] & &\\
    \hline
    2024/10/25 &02:25:51 &0.034 &19.03$\pm$0.12 &89$^{+10}_{-9}$ &TRT-SRO &\cite{Jiang2024GCN}\\
    2024/10/25 &02:37:00 &0.042 &18.97$\pm$0.04 &94$^{+4}_{-3}$ &COLIBRÍ-VIS &\cite{Watson2024GCN}\\
    2024/10/25 &02:39:42 &0.044 &18.90$\pm$0.46 &100$^{+52}_{-34}$ &KNC &This work\\
    2024/10/25 &03:48:09 &0.091 &18.90$\pm$0.16 &99$^{+16}_{-13}$ &KNC &This work\\
    2024/10/25 &04:58:39 &0.14 &19.21$\pm$0.05 &75$^{+4}_{-3}$ &TRT-SRO &This work\\
    2024/10/25 &05:31:41 &0.16 &19.61$\pm$0.24 &52$^{+12}_{-10}$ &KNC &This work\\
    2024/10/25 &08:04:33 &0.27 &19.78$\pm$0.08 &44$^{+3}_{-3}$ &TRT-SRO &This work\\
    2024/10/25 &12:12:03 &0.44 &20.47$\pm$0.08 &24$^{+2}_{-2}$ &TNOT &This work\\
    2024/10/25 &14:12:49 &0.52 &20.37$\pm$0.08 &26$^{+2}_{-2}$ &TNOT &This work\\
    2024/10/25 &17:18:57 &0.65 &21.37$\pm$0.35 &10$^{+4}_{-3}$ &RC80 &\cite{Vinko2024GCN}\\
    2024/10/26 &02:58:26 &1.06 &20.60$\pm$0.19 &21$^{+4}_{-3}$ &TRT-SRO &This work\\
    2024/10/28 &04:58:26 &3.14 &22.9$\pm$0.1 &2.4$^{+0.2}_{-0.2}$ &LBT &\cite{Rossi2024GCN}\\
    \hline
    \end{tabular}}
    \tablefoot{Uncertainties are reported with a $1\sigma$ confidence level. Magnitudes (fourth column) and flux densities (fifth column) are corrected for the Galactic extinction using the \texttt{dust\_extinction} Python package \citep{Gordon2024} with the F19 model for the Milky Way, R$_V=3.1$ and $E(B-V)=0.249$ \citep{Klingler2024GCN, Schlegel1998}.}
    \label{tab:optical_data}
\end{table}

\begin{table}
    \centering
    \caption{Log table of our $I$-band and $i$-band observations and publicly available data of GRB\,241025A.}
    \resizebox{\columnwidth}{!}{
    \begin{tabular}{lcccccc}
    \hline
     Date & UTC  &T$_m$ &Mag &$F_{\nu}$ &Instrument &Ref\\
          &[hh:mm UT] &[days] &[AB] &[$\upmu$Jy] & &\\
    \hline
    2024-10-25 &06:35:00 &0.21 &18.88$\pm$0.08 &102$^{+8}_{-7}$ &TRT-SRO &This work\\
    2024-10-25 &08:25:50 &0.28 &19.05$\pm$0.16 &87$^{+14}_{-12}$ &TRT-SRO &This work\\
    2024-10-25 &16:08:58 &0.60 &19.79$\pm$0.08 &44$^{+3}_{-3}$ &GIT &\cite{Mohan2024GCN}\\
    2024-10-25 &17:18:57 &0.65 &19.56$\pm$0.17 &54$^{+9}_{-8}$ &RC80 &\cite{Vinko2024GCN}\\
    2024-10-26 &02:57:24 &1.06 &20.04$\pm$0.47 &35$^{+19}_{-12}$ &KNC &This work\\
    2024-10-26 &03:29:46 &1.08 &20.15$\pm$0.28 &32$^{+9}_{-7}$ &TRT-SRO &This work\\
    \hline
    \end{tabular}}
    \tablefoot{Uncertainties are reported with a $1\sigma$ confidence level. Magnitudes (fourth column) and flux densities (fifth column) are corrected for the Galactic extinction using the \texttt{dust\_extinction} Python package \citep{Gordon2024} with the F19 model for the Milky Way, R$_V=3.1$ and $E(B-V)=0.249$ \citep{Klingler2024GCN, Schlegel1998}.}
    \label{tab:NIR_data}
\end{table}

\begin{table}
    \centering
    \caption{Log of EP/FXT observations and results of the spectral fits.}
    \resizebox{\columnwidth}{!}{
    \begin{tabular}{lccccc}
    \hline
     Date &UTC & Exposure  & $N_{\mathrm{H,int}}$ & $\mathrm{log}_{10}F$ & $\alpha$ \\
          &[hh:mm:ss] UT &[sec]     & [10$^{22}$\,cm$^{-2}$]   & [erg cm$^{-2}$ s$^{-1}$] & \\
    \hline
      2024-10-25 &09:34:17.906 & 2873       & 4.37$^{2.78}_{-2.54}$    & -11.38$_{-0.05}^{+0.05}$ & 1.77$^{+0.21}_{-0.21}$ \\ 
      2024-10-26 &14:27:59.892 & 2769      & 4.37$^{2.78}_{-2.54}$    & -12.56$_{-0.19}^{+0.21}$ & 3.11$^{+0.93}_{-0.73}$ \\ 
      2024-10-27 &08:02:52.425 & 2985    & 4.37$^{2.78}_{-2.54}$    & -12.82$_{-0.44}^{+0.25}$ & 3.11$^{+0.93}_{-0.73}$ \\ 
    \hline
    \end{tabular}}
    \tablefoot{The parameters of the fits are reported with the errors representing the 95\% confidence interval. $N_{\mathrm{H,int}}$ is the intrinsic host galaxy absorption; $F$ is the unabsorbed flux integrated between 0.3--10\,keV; $\alpha$ is the photon index.}
    \label{tab:FXT}
\end{table}

\section{Scintillation effects}
\label{appendix:scintillation}
\begin{figure*}
\centering
    \includegraphics[width=0.85\textwidth]{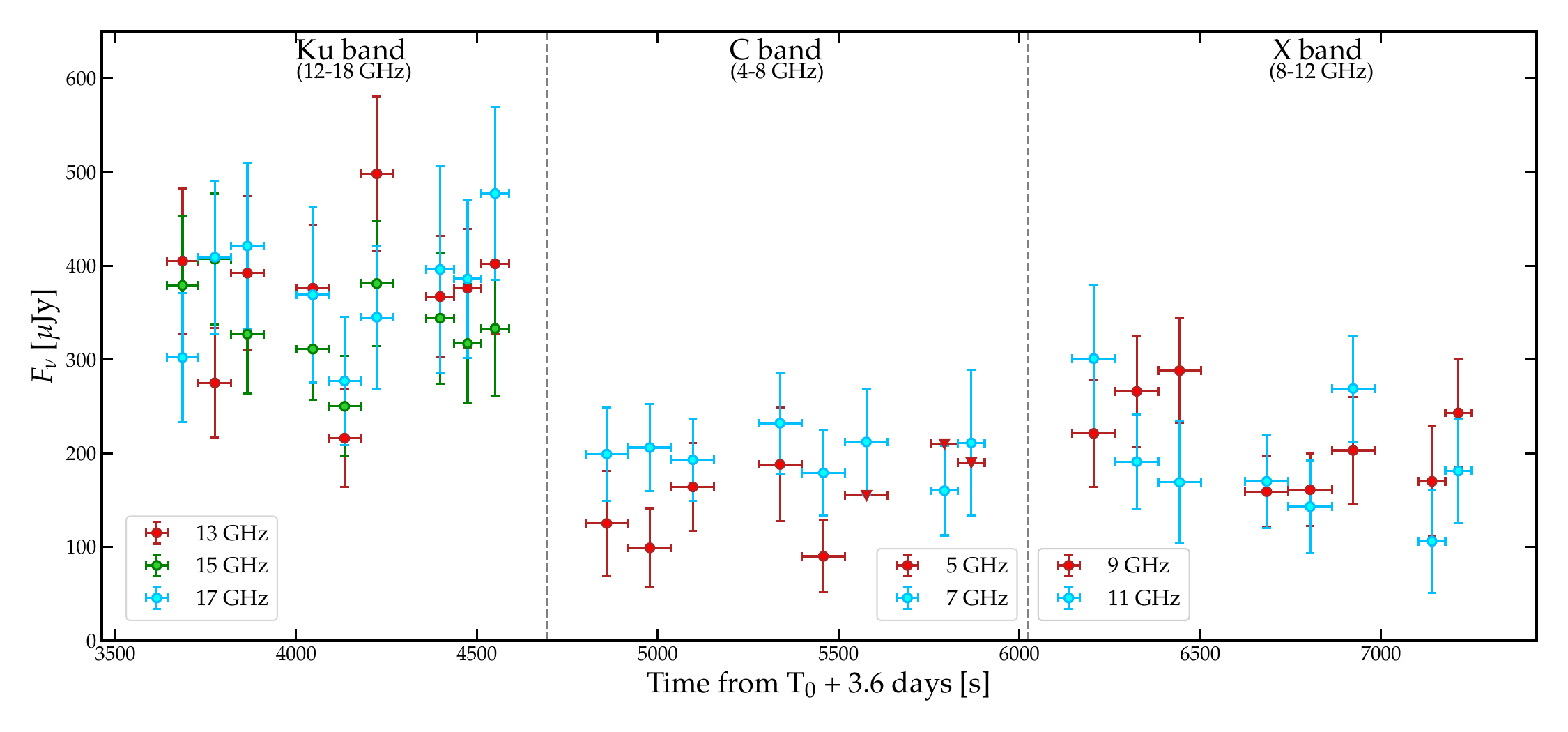}
    \caption{Spectrum evolution at 3.65 days post-burst. Each sub-band (see Table\,\ref{tab:grb25A_radio_log}) has been split in 1-2\,min chunks, according to the duration of each scan. Upper limits are reported at a 5$\sigma$ confidence level, given that the overall quality of each image is significantly poorer with respect to the data presented in Table\,\ref{tab:grb25A_radio_log}.}
    \label{fig:scintillation}
\end{figure*}

Radio emission from the afterglow phase of GRBs can suffer from scintillation effects as it propagates through the turbulent ionised ISM of the Milky Way. Interstellar scintillation can produce significant fluctuations in the observed flux density that are frequency dependent and can last from minutes to days post-burst \citep{Rickett1990, Goodman1997, Walker1998, Chandra2008, Alexander2019}. 
In this appendix we investigate whether our observations may be affected by interstellar scintillation. The radio light curves and spectra were presented in Figure\,\ref{fig:semi-empirical-lc} and Figure\,\ref{fig:semi-empirical-spec}, respectively. The light curves at each frequency do not show evidence for extreme variability ($\sim$an order of magnitude, or more, jumps in flux density from an epoch to another), namely the observed trends can be traced back to the intrinsic afterglow evolution. The spectra at 0.76, 7.51, 15.47, 30.59 and 67.43 days post-burst do not show any evidence of abrupt (i.e., an order of magnitude) changes in the flux density between different frequency bands.

Conversely, the spectrum at 3.65 days shows a significant ($>3\sigma$) step-like increase between the flux density estimate at X-band (9 and 11\,GHz) and the one at Ku-band (13, 15 and 17\,GHz). As diffractive scintillation might affect the data at these early times, we further split the data of this epoch in time (1-2 minute-long chunks for each scan, depending on the total duration of each scan). Results are shown in Figure \ref{fig:scintillation}. We do not find any significant ($\geq$3$\sigma$) variability between different frequency bands, nor within individual frequency bands, which cannot be traced back to the intrinsic afterglow evolution. In fact, at this epoch all the radio bands are below the maximum of the spectrum, and hence $F_{4-8\,\mathrm{GHz}} \leq F_{8-12\,\mathrm{GHz}} \leq F_{12-18\,\mathrm{GHz}}$. In conclusion, we do not find any evidence of scintillation in our observations.

\section{X-ray fit}
\label{appendix:butterfly}
\begin{figure}
\centering
    \includegraphics[width=0.65\columnwidth]{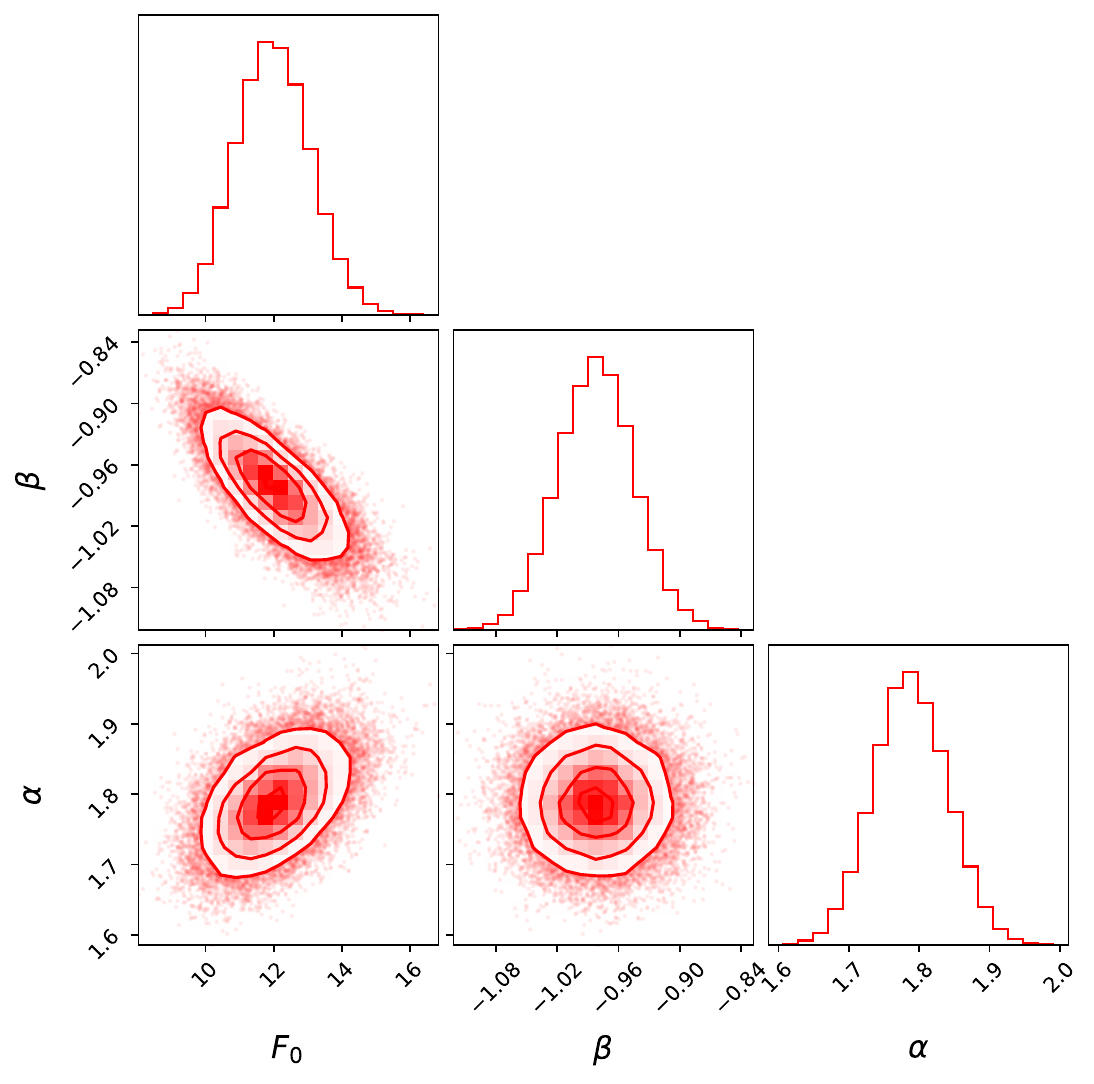}
    \includegraphics[width=0.65\columnwidth]{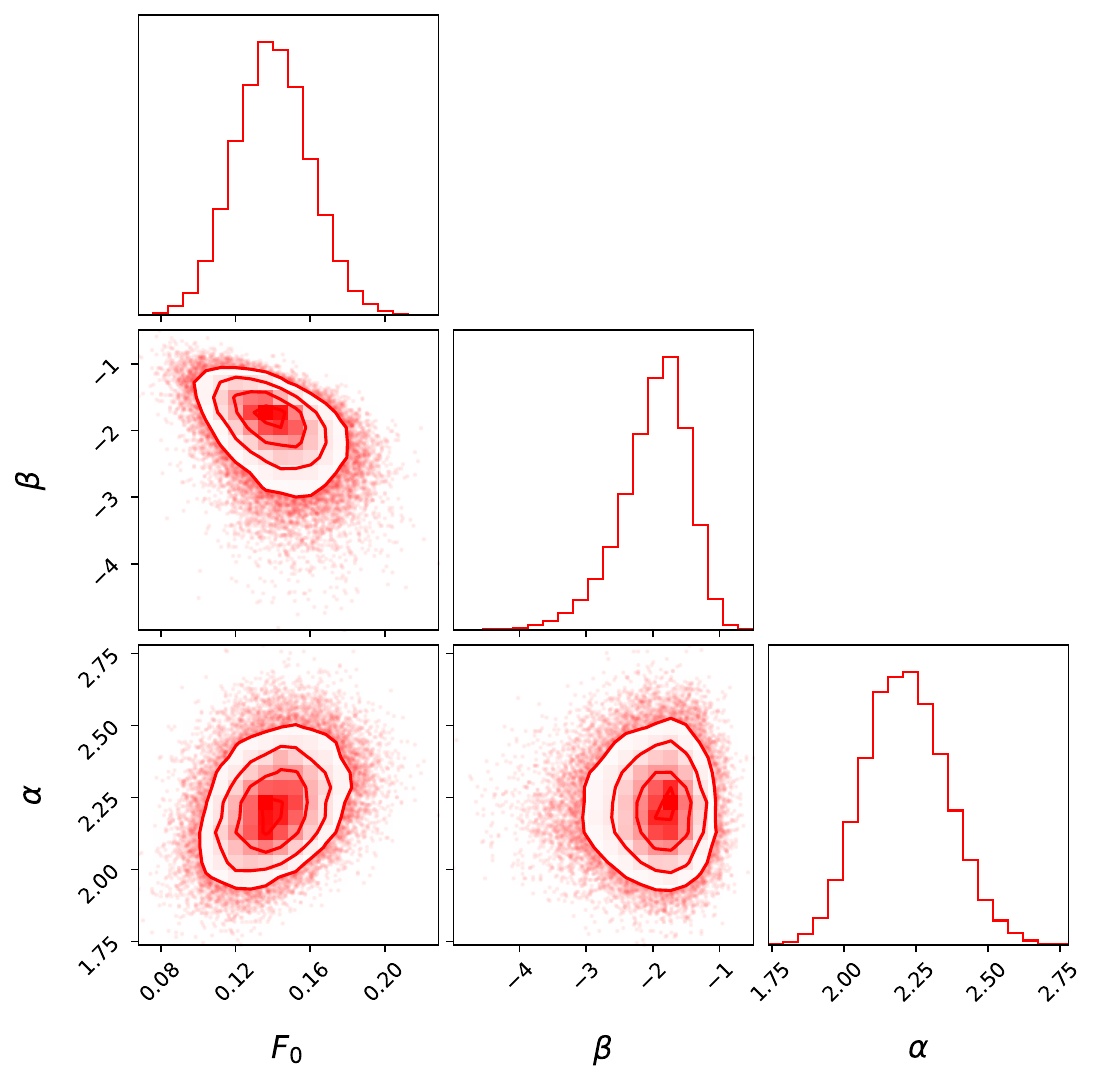}
    \caption{Corner plot of the parameters obtained from the fit of the \textit{Swift}/XRT and EP/FXT integrated flux measurements before (upper panel) and after (lower panel) $t=7\times10^4$\,s post-burst. $F_0$ is the normalisation at 1\,keV in $\upmu$Jy, $\beta$ is the temporal index and $\alpha$ is the photon index.}
    \label{fig:corner_Xrays}
\end{figure}

\begin{figure}
\centering
    \includegraphics[width=0.6\columnwidth]{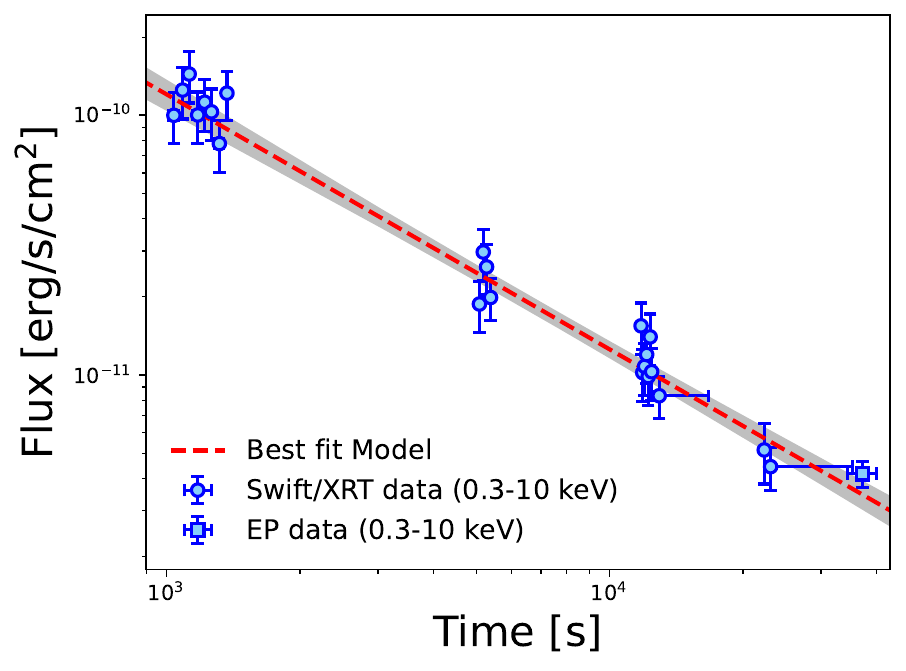}
    \includegraphics[width=0.63\columnwidth]{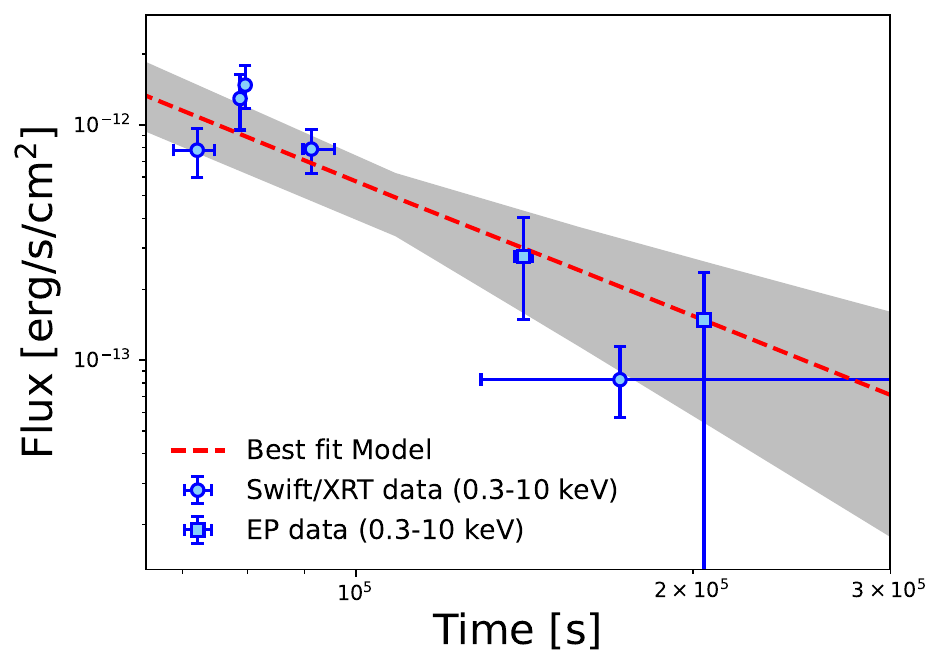}
    \caption{Fit of the X-ray integrated flux measurements before (upper panel) and after (lower panel) $t=7\times10^4$\,s post-burst. Data are presented as blue dots (\textit{Swift}/XRT) and blue squares (EP/FXT). The dashed red line and the gray region represent the median and the 90\% credible interval of the posterior distribution, respectively. The last \textit{Swift}/XRT point in the right panel was excluded from the fit (see the text).}
    \label{fig:fit_Xray_break}
\end{figure}

In this appendix we report the fit of the X-ray data before and after the break at $\sim 7\times10^4$\,s post-burst. First, we assumed that the flux density can be parametrised as $F_{\nu} = F_0t^{\beta}\nu^{1-\alpha}$, where the three parameters are $F_0$, i.e. the normalisation at 1\,keV in $\upmu$Jy, $\beta$ and $\alpha$, namely the temporal and photon indices, respectively. To fit the unabsorbed X-ray flux, we built the fitting curve by integrating $F_{\nu}$ over the 0.3 -- 10\,keV energy range. The fit was then performed on both the integrated flux measurements and the photon indices provided by the SWIFT BURST ANALYZER (for \textit{Swift}/XRT) and Table\,\ref{tab:FXT} (for EP/FXT) within a Bayesian framework, assuming uniform prior distributions for the three parameters and sampling the posterior probability distributions with the MCMC algorithm implemented in the \texttt{emcee} python package \citep{Foreman-Mackey2013}. Figure \ref{fig:corner_Xrays} shows the corner plot of the three parameters for the fit of the data pre-break (upper panel) and post-break (lower panel). Table \ref{tab:pre-break-fit} reports the median and the 90\% uncertainty on the posterior distribution of each parameter. We note that the last \textit{Swift}/XRT data point combines data from two different observations. Since the photons within this bin are not uniformly distributed in time, the computed central time of the bin is not reliable. Given that our fitting procedure strongly depends on this central temporal value of the bin, we excluded the last \textit{Swift}/XRT point from the fit. Finally, Figure \ref{fig:fit_Xray_break} shows the X-ray light curve and the modelling curve. 
Once the posterior distributions of the normalisation $F_0$, the photon index $\alpha$ and the temporal index $\beta$ were derived, we computed the distribution of $F_{\nu}$ between 0.3 and 10\,keV at each epoch by simply fixing the time $t$ of that epoch. The butterfly plots on the right panel of Figure \ref{fig:LC_and_spectra} were then derived by considering the 90\% confidence interval of the $F_{\nu}$ distribution at a given epoch.
\begin{table}
    \centering
    \caption{Median and 90\% uncertainty on the parameters derived with the fit on \textit{Swift}/XRT and EP/FXT data before (first row) and after (second row) $t=7\times10^4$\,s  post-burst.}
    \begin{tabular}{cccc}
    \hline
    $F_0$ &$\alpha$ &$\beta$ &Time range of the fit \\
    \text{[$\upmu$Jy]} & & &[sec post-burst]\\
    \hline
    $12^{+2}_{-2}$ &$1.79^{+0.09}_{-0.08}$ &$-0.98^{+0.06}_{-0.06}$ &$<7\times10^4$\\
    $0.14^{+0.03}_{-0.03}$ &$2.2^{+0.2}_{-0.2}$ &$-1.9^{+0.6}_{-1.0}$ &$>7\times10^4$ \\ 
    \hline
    \end{tabular}
    \tablefoot{The normalisation $F_0$ is computed at 1\,keV, at the time of the first X-ray observation considered for the fit. We excluded the last \textit{Swift}/XRT point from the fit (see the text).}
    \label{tab:pre-break-fit}
\end{table}

\section{Semi-empirical radio evolution model corner plot}
\label{appendix:semi-empirical}
\begin{figure*}
    \centering
    \includegraphics[width=\textwidth]{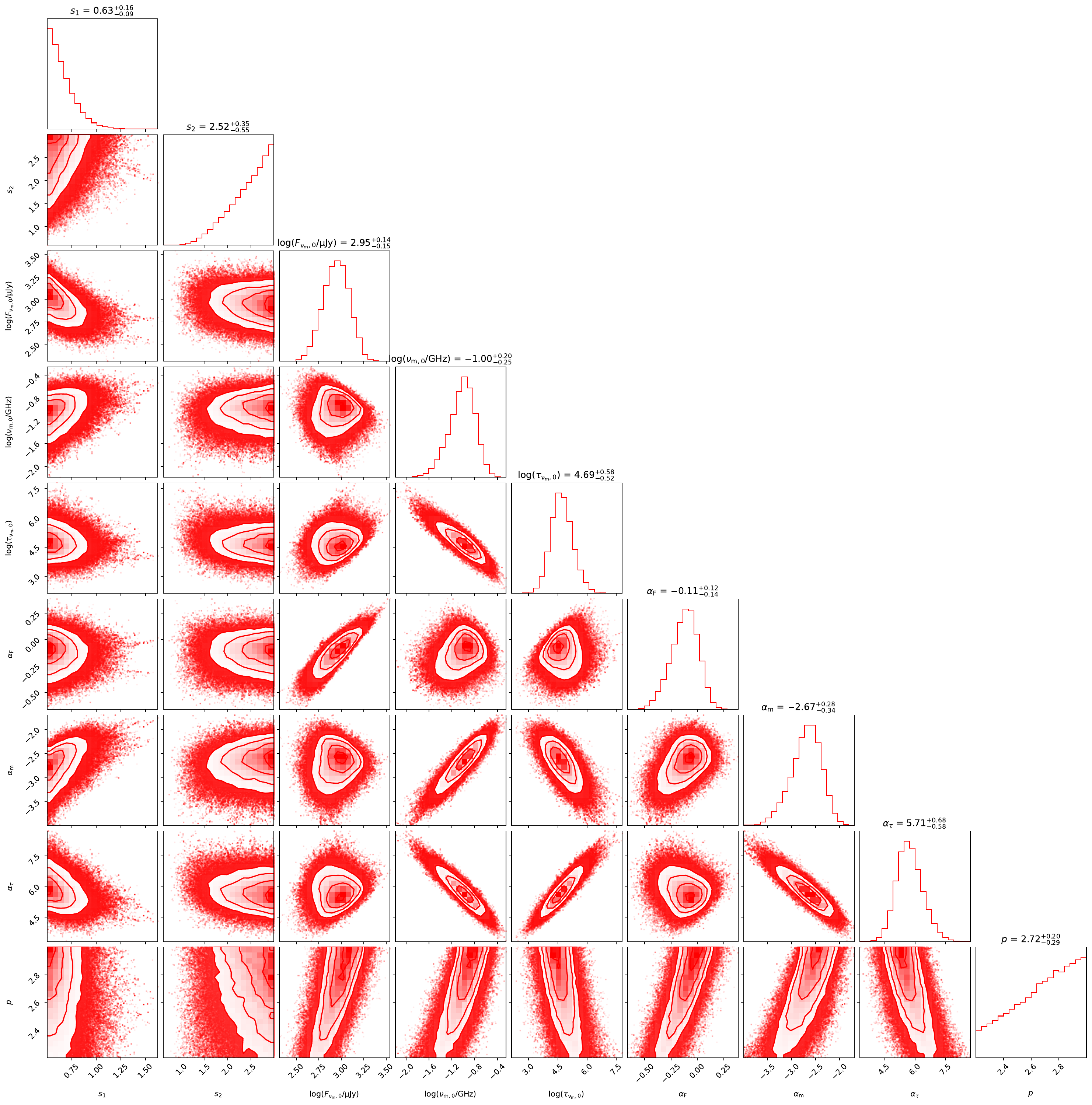}
    \caption{Corner plot of the semi-empirical model (cfr.\ Sect. \ref{sec:semi_emp_radio}) posterior probability. In the panels under the diagonal, the contours show the one-, two- and three-sigma credible regions of the marginalized two-dimensional posteriors. Dots show samples outside the three-sigma contour. The panels on the diagonal show the marginalised one-dimensional posterior probability distribution of each single parameter, with the median and 68\% credible interval shown on top of the panel.}
    \label{fig:semi_emp_corner}
\end{figure*}

In Figure \ref{fig:semi_emp_corner} we show a corner plot illustrating the posterior probability distribution of the parameters of the semi-empirical model of the radio evolution described in Section \ref{sec:semi_emp_radio}.

\section{Standard afterglow model parameter exploration}
\label{appendix:model_param_exploration}
\begin{figure*}
    \centering \includegraphics[width=\textwidth]{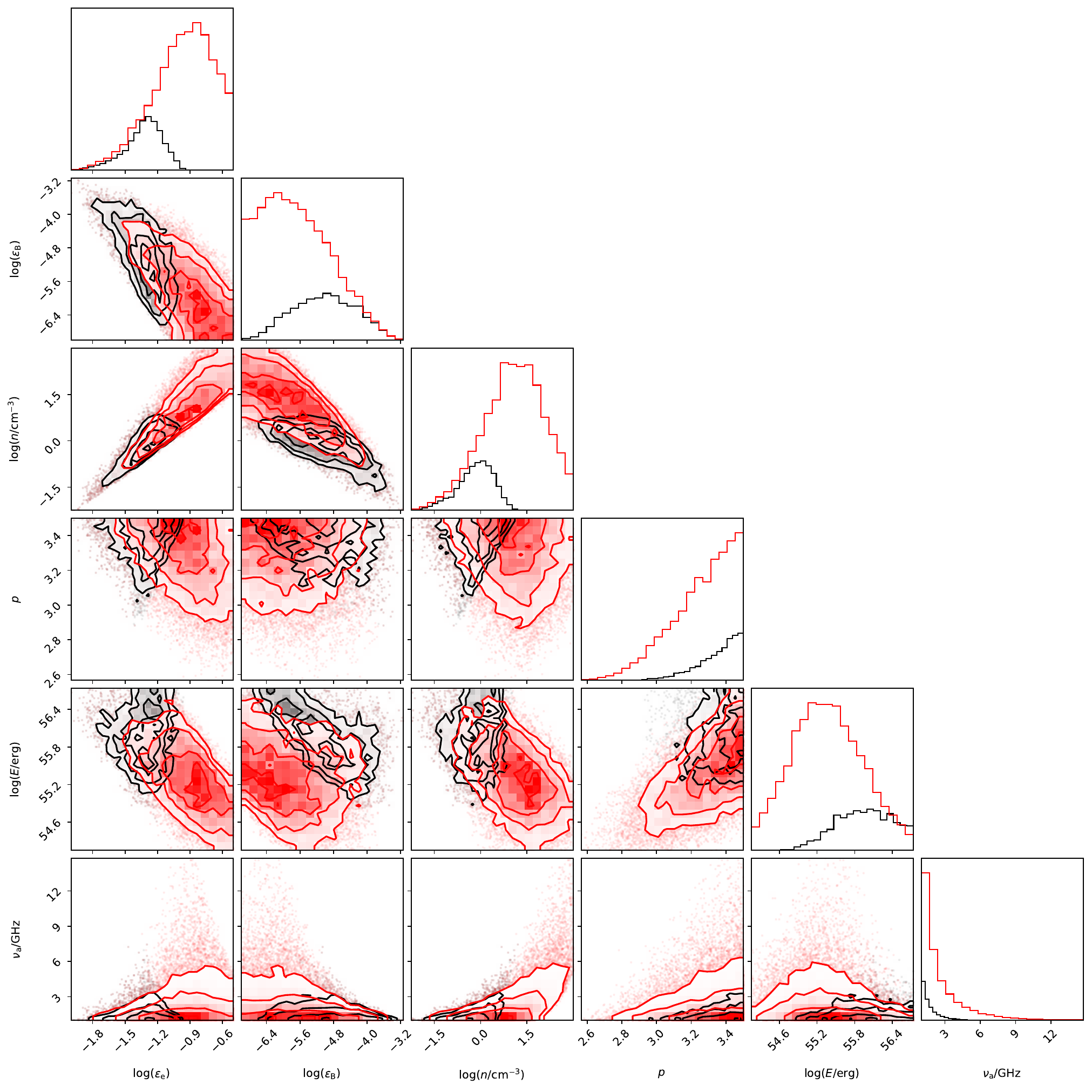}
    \caption{Corner plot showing the distribution of samples that survive the cuts that represent the constraints imposed by the positions of the break frequencies and the X-ray flux density during the self-similar phase, as discussed in Sect.\ \ref{sec:standard_afg_model_failure}. The black contours and histograms refer to the model that includes SSC cooling in the Thomson regime (following \citealt{Sari2001}). The red contours and histograms refer to a model that ignores SSC.}
    \label{fig:params_space_corner}
\end{figure*}

The corner plot in Figure \ref{fig:params_space_corner} demonstrates the impact of the constraints discussed in Sect.\ \ref{sec:standard_afg_model_failure} on the parameter space of a standard afterglow model. The results including SSC in the Thomson regime are shown in black, while those obtained neglecting SSC cooling are shown in red.

\section{Multi-wavelength afterglow model without the enhanced self-absorption optical depth}
\label{appendix:model_wo_enhancement}
\begin{figure*}[h!]
    \centering
    \includegraphics[width=0.95\textwidth]{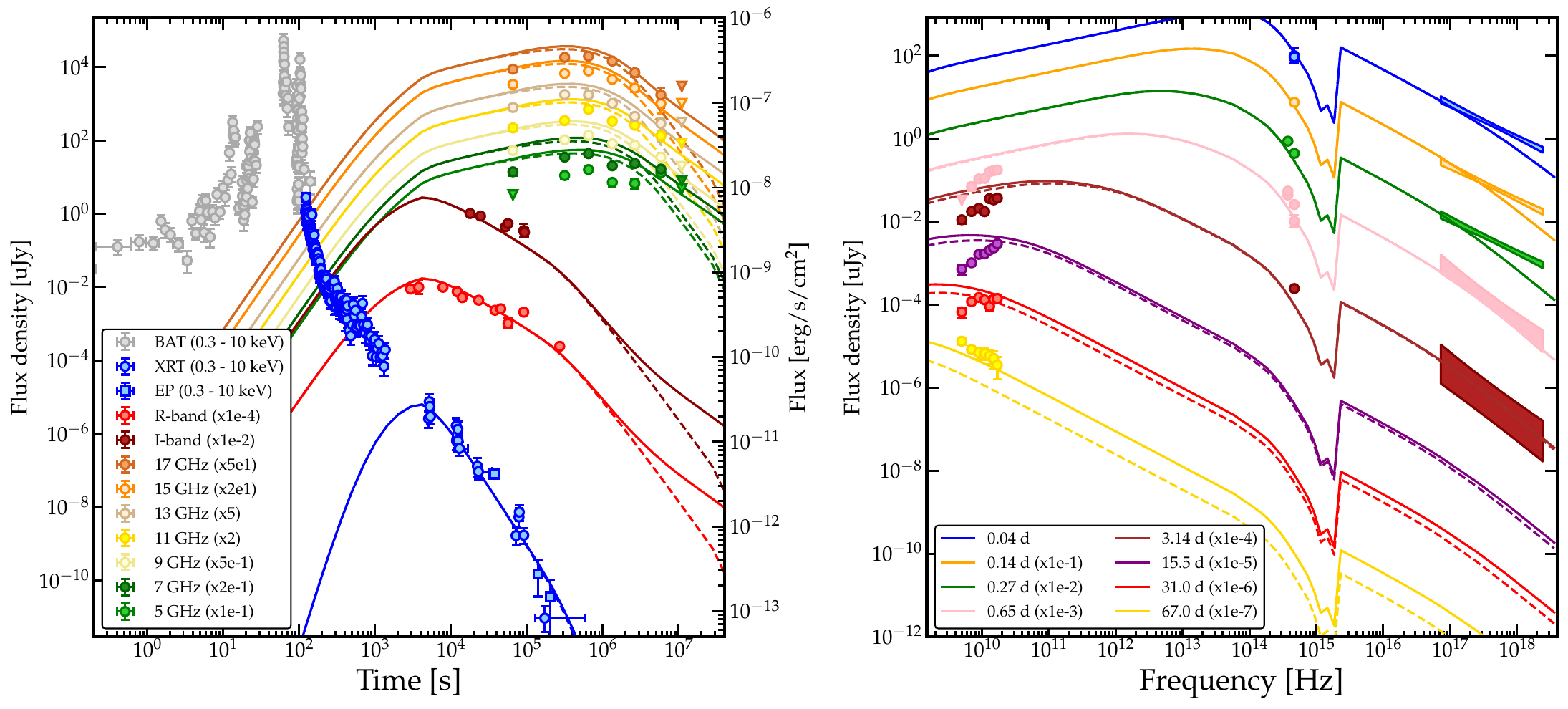}
    \caption{Observations and afterglow modelling of GRB\,241025A. \textit{Left panel}: multi-wavelength light curves. \textit{Right panel}: spectra at different times. The afterglow model, represented with dashed (core of the jet) and solid (total emission, i.e. core + wings) lines, is computed with the set of parameters of Table \ref{tab:model_params} but $\tau_{\mathrm{enh}}$, which is set to one.}
    \label{fig:model_wo_enhancement}
\end{figure*}
In Figure\,\ref{fig:model_wo_enhancement} we show the equivalent afterglow modelling presented in Figure\,\ref{fig:LC_and_spectra}, without enhanced optical depth. The modelling light curves and spectra were computed with the same parameter choice of Table \ref{tab:model_params}, except $\tau_{\mathrm{enh}}$, which is set to $1$. It is clear that the colour evolution in radio cannot be reproduced without artificially increasing the optical depth. 

\end{appendix}

\end{document}